\documentclass[11pt,a4paper]{article}
\usepackage[latin1]{inputenc}
\usepackage{amsmath}
\usepackage{amsfonts}
\usepackage{amssymb}
\newcommand{\s}{\sigma }

\newcommand{\hf}{\frac{1}{2}}

\newcommand{\nb}{{\bar {n}}}
\newcommand{\mb}{{\bar {m}}}
\newcommand{\xn}{x_{n}}
\newcommand{\xnb}{{\bar x _{n}}}

\newcommand{\xm}{x_{m}}
\newcommand{\xmb}{\bar x _{m}}

\newcommand{\e}{e^{i k_{0}Y}}                                      

\newcommand{\kim}{ k_{1\mu}}                                      
\newcommand{\kom}{ k_{0\mu}}                                      
\newcommand{\ki}{ k_{1}}
\newcommand{\yn}{ Y_{n}}                                             
 
\newcommand{\kn}{ k_{n}}

\newcommand{\kt}{ k_{2}}                                             
\newcommand{\ko}{ k_{0}}                                             
\newcommand{\yim}{ Y_{1}^{\mu}}                                      
\newcommand{\yin}{ Y_{1}^{\nu}}                                      
\newcommand{\kin}{ k_{1\nu}}  
                                    
\newcommand{\kon}{ k_{0\nu}}
\newcommand{\kor}{ k_{0\rho}}                                      
\newcommand{\ktm}{ k_{2\mu}}   
                                    
\newcommand{\ytm}{ Y_{2}^{\mu}}                                      
                                                              
\newcommand{\lpp}{\mbox {$e^{i\int _{c} \alpha (t)                             
k(t) \partial _{z} X(z+t) dt +ik_{0}X}$}}

\newcommand{\gvk}{ e^{i\sum _{n }k_{n}Y_{n}}}

\newcommand{\p}{\partial}                                           
\newcommand{\pp}{\partial ^{2}}

\newcommand{\li}{ \lambda_{1}}                                    
\newcommand{\lt}{ \lambda_{2}}                                    
\newcommand{\eps}{ \epsilon}                                        
\newcommand{\al}{\alpha }                                             
\newcommand{\aln}{\alpha _{n}} 
                                 
\newcommand{\lan}{\langle}
\newcommand{\ran}{\rangle}

\newcommand{\xb}{\mbox{$\bar{x}$}}

\newcommand{\zb}{{\bar{z}}}                                             
\newcommand{\tb}{\mbox{$\bar{t}$}}

\newcommand{\kib}{\mbox {$ \bar{k_{1}}$}}

\newcommand{\la}{ \lambda }                                           
\newcommand{\be}{\begin{equation}}                                             
\newcommand{\br}{\begin{eqnarray}}                                             
\newcommand{\ee}{\end{equation}}                                               
\newcommand{\er}{\end{eqnarray}}

\renewcommand{\theequation}{\thesubsection.\arabic{equation}}

\begin{document}
\title{
\hfill\parbox{4cm}{\normalsize IMSC/2013/05/04\\
}\\
\vspace{2cm}
Loop Variables and Gauge Invariant Exact Renormalization Group
Equations for Closed String Theory 
%\thanks{}%
}
\author{B. Sathiapalan\\ {\em                                                  
Institute of Mathematical Sciences}\\{\em Taramani                     
}\\{\em Chennai, India 600113}}                                     
\maketitle                                                                     
\begin{abstract}   
We formulate the Exact Renormalization Group on the string world sheet for closed string backgrounds.  The same techniques that were used for open strings is used here. There are some subtleties. One is that holomorphic factorization of the closed string vertex operators does not hold in the presence of a cutoff on the Euclidean world sheet. This introduces
extra terms in the Lagrangian at the cutoff scale and they turn out to be crucial for implementing gauge invariance. This naive generalization from open string to closed strings requires a {\em massive} graviton and the gauge symmetry is Abelian, just as in open string theory. Interestingly, it turns out that if one introduces a non dynamical background metric (as in background field formalism) and combines a gauge transformation on the field with a transformation on the coordinates and background metric,  the graviton can be massless.  Some examples of  background coordinate covariant equations are worked out explicitly. A preliminary discussion of massive modes, massive gauge transformations and  the role of world sheet regulator terms is given. Some
of the gauge transformations can be given a geometric meaning if space time is assumed to be complex at some level.

 \end{abstract}                                                                 
\newpage                                                                       
\section{Introduction} 

Equations of Motion(EOM) for specific open or closed string modes can be obtained as Renormalization Group (RG) equations (i.e. $\beta$-functions) for the world sheet action of a string propagating in a non trivial background [\cite{L}-\cite{T}]. The Exact Renormalization Group (ERG) \cite{WK,W,W2,P} on the other hand, would include all the modes of the string at once and be equivalent to string field theory as first suggested in \cite{BM,HLP}. Some aspects of this and the connection with the proper time formalism was worked out in \cite{BSPT}. 

String theory possesses an infinite tower of gauge symmetries and one would like the EOM to be gauge invariant. Ideally one would like an action too. 
In string field theory this was elegantly solved using the BRST formalism \cite{SZ,WS,Wi2,BZ} and an action was written down.

The problem of gauge invariance can also be posed in the RG formalism. The RG formalism is potentially capable of being manifestly background independent, so a solution to the problem of gauge invariance should provide insights into fundamental aspects of string theory.
In the RG formalism, loop variable techniques have been used to make  equations gauge invariant \cite{BSLV} in the free case.
Interacting RG equations were also made gauge invariant \cite{BSREV} though not in a form that is conveniently written down
in terms of space-time fields. A convenient form was derived more recently
in \cite{BSERGopen1} and \cite{BSERGopen2}, (hereafter I and II) where gauge invariant interacting equations of motion for open strings were derived. This was obtained by applying the ERG to the world sheet action for open strings propagating in an arbitrary background. Loop variable techniques were used to ensure that the equations are gauge invariant. Since the world sheet action can be written for any background, this method is manifestly background independent \footnote{In the BRST formalsim background independence has been discussed in \cite{Wi,LW,Sh,KMM}}.  The equations are quadratic, as expected
from open string field theory. This can be traced to the fact that the ERG is always quadratic in coupling constants. Thus if equations of motion can be obtained from an ERG they are guaranteed to be quadratic. 

	The unexpected feature of the equations is that the interactions between all modes, including massive ones, are in the form of gauge invariant "field strengths" just like the Dirac-Born-Infeld equations for the massless sector in open string theory. Furthermore the gauge transformations are of the same form as in the free theory. This is characteristic of an Abelian theory. In the absence of Chan-Paton factors, open string gauge invariance is, in fact Abelian. In the BRST formulation, however, this is not the case. It is possible that a match between the two can be achieved with some field redefinitions.
	
	Another interesting feature was that the equations seem to have their origin in a massless theory in one higher dimension. This pattern was verified in some detail for the first few levels and depended on the existence of solutions
to a highly over complete set of linear (algebraic) equations and, a priori, is quite non trivial. Further insight is required to understand
why this is the case.

A natural question is to enquire whether any of this can carry over to closed strings. In particular if a gauge invariant ERG can be written, it is guaranteed to be quadratic. Of course the basic interaction vertices in closed string theory are also cubic. We also know that the OPE (operator product expansion) of string vertex operators carries all the information about string interactions. From this point of view it should be possible to write a cubic action and thus, quadratic equations for closed string also.  In the RG approach, one expects that
the ERG is quadratic in coupling constants. It is only after solving for all the irrelevant couplings that the full non polynomial $\beta$ function for the marginal coupling emerges.

However in BRST closed string field theory the action is non polynomial \cite{BZ}. It is at first sight surprising that the main conclusion of this paper is that indeed gauge invariant quadratic equations can be written down starting from an ERG of the world sheet theory for closed strings, just as in open string theory. However the gauge invariance here involves transforming the background field. This is not 
the same as the original invariance. Nevertheless (as in usual background field formalism) because it is very similar to the full symmetry it is useful to have manifest at intermediate stages of the calculation.

The technique used for open strings in I and II, continues to be applicable here with one new ingredient. In closed string theory the world sheet equation $ \p _ z \p _{\bar z} X(z,\bar z)=0$ ensures that the vertex operators are all of the form $\p_z^n X \p_{\bar z}^mX$ and do not involve any mixed derivatives: $\p_z ^n\p_{\bar z}^m X$. This can also be seen from the fact that Green function (in the plane) \[
G(z,\bar z;0)=\lan X(z,\bar z) X(0) \ran = ln~ (z \bar z) = ln~z + ln ~\bar z =\lan X(z) X(0)\ran+\lan \bar X(\bar z) \bar X(0) \ran\] breaks up into a holomorphic and anti holomorphic part.
However in the presence of a world sheet cutoff this is not true in general. For instance a short distance cutoff Green function would be $G(z,\bar z;0;a) = ln~(|z|^2 + a^2)$, which does not break up into a holomorphic and anti holomorphic part.\footnote{If we use a Lorentzian world sheet metric, it is possible to regulate the left and right sectors separately: $G(x_L,x_R;0) = ln~ (x_L^2+a^2) + ln~(x_R^2+a^2)$. Since $x_L , x_R $ are real, this is a valid regularization.}

This suggests that at least away from the continuum limit, when a finite cutoff is present, one should have vertex operators involving mixed derivatives. Indeed we will see below that it is essential for gauge invariance. These vertex
operators will not contribute to the S-matrix because their correlators vanish in the continuum limit by using the equations of motion. 

The final result is very similar to that of open strings: We get quadratic equations of motion with interactions in terms of gauge invariant "field strengths" and gauge transformation law {\em unmodified by interactions}. This last fact is difficult to reconcile with what we know about gauge transformations in gravity: general coordinate transformations (GCT), which are definitely non-Abelian in form. 

At first sight, another problematic feature in this construction is that the field strength construction for the lowest mass state, viz the graviton, makes sense only if the graviton is {\em massive}. There is a vector field at this level coming from the mixed (non-holomorphic) vertex operator which, in the unitary gauge is "eaten up" by the graviton, (much as a Goldstone boson is eaten up by a vector field). 

Both these problems are resolved by modifying the gauge transformation by including transformations of the coordinates.
The first problem (i.e. that the transformation is Abelian) is solved because this induces a tensor rotation of the fields (since they multiply vertex operators).
\footnote{This somewhat similar to what was suggested in \cite{BSCP} where in the presence of Chan Paton factors, there is an extra group rotation symmetry, which  give rise to the non Abelian rotation term.} The second problem is solved because the kinetic term involving $\eta_{\mu \nu}\p_z X^\mu \p_\zb X^\nu$ is not invariant, and making it invariant involves introducing a background metric and induces an extra term in the definition of the field strength. This non dynamical term (essentially a Christoffel connection for the non dynamical background metric) plays the role of the extra vector field, which is not needed anymore. The graviton thus remains massless. 

At this point one makes contact with standard general relativity. The massive graviton phase is analogous to the unbroken phase of a scalar field theory (where the scalar field is the graviton). The massless phase is analogous to the Goldstone phase where $\lan g_{\mu \nu} \ran = \eta _{\mu \nu}$. The equations for the massless graviton is covariant under  {\em background} GCT with a background metric. The original gauge transformation is now part of this symmetry.

This idea can be transcribed in terms of loop variables, which is required for gauge invariance of massive higher spin fields \cite{F,SH} .  Massive field equations can be written down. One has to suitably define general coordinate transformations such that these massive fields are tensors. If one assumes this, it is easy to write down generally covariant equations. They are also invariant under the (massive) gauge transformations independently. \footnote{Note that the gauge transformation of the (massless) graviton is tied to the background coordinate transformations, but the massive gauge transformations are independent.} Thus when all the dust settles, we  have gauge invariant and background generally covariant equations for all modes. There is also possibly a space time interpretation for massive gauge transformation. This however requires further analysis.

This paper is organized as follows: In Section 2 we give a brief summary of the techniques used in I and II and the generalization of this technique to closed strings. In Section 3 we give some explicit calculations, for the lowest and second mass levels. In Section 4 we discuss the connection of the gauge transformations described here and general coordinate transformations. Section 5 discusses the massive modes.  Section 6 contains a summary and conclusions. 

\section{Background}

We assume the background material in I,II and earlier papers, suitably generalized for the present discussion. One difference in notation is that we have consistently used lower indices for $k_\mu (t)$ and upper indices for $X^\mu$. Index contractions can be done using $\eta_{\mu \nu}$ because we are in flat space. But in Section 4 we will introduce a background (but still flat) metric $g_{\mu \nu}^R$ and then one has to be more careful. Typically contractions will be done using the background metric unless otherwise specified.
\subsection{ERG}

We first write down an ERG in position space. \footnote{For discussions on various aspects of ERG see \cite{BB1,BB2,S1,IIS} in addition to the original references. The position space ERG for the string world sheet has also been discussed in \cite{HLP,BSERG,BSERGTach}.}  
We start with a Euclidean field theory on the world sheet. $z$ describes the world sheet coordinates. For open strings,
$z=x$ is on the $x$-axis. For closed strings $z=x+i y$. Thus for closed strings $\int dz$ should be understood as $\int d^2z$ and $X(z) = X(z,\zb)$. 
 The action is:
\[
S= \underbrace{-\hf \int dz ~\int dz' ~Y^\mu(z) (G^{-1})_{\mu\nu}(z,z';\tau)Y^\nu(z')}_{Kinetic~ term} + \underbrace{\int dz~L[Y^\mu(z),Y^\mu_{n,\bar m}(z)]}_{Interaction}
\]
Here, $G^{\mu\nu}(z,z';\tau)\equiv\lan Y^\mu(z) Y^\nu(z')\ran$ is a cutoff propagator, where $\tau$ parametrizes the cutoff. Thus for instance we can take $\tau=~ln~a$ where $a$ is a short distance cutoff or lattice spacing. $\mu =0,....,D-1$ are the usual space time coordinates and $\mu= D$ (we refer to the coordinate as $\theta$), is the coordinate that plays the role of the bosonized ghost of the BRST formalism. We will take $G^{\mu\nu}(z,z';\tau)=\eta^{\mu \nu}G(z,z';\tau)$ for $\mu=0,...,D-1$ and $G^{DD}(z,z';\tau)=\lan \theta (z)\theta (z')\ran$. We take $\theta$ to be a {\em massive} field, with a mass of the order of the cutoff $m=O({1\over a})$. This choice is made in order to reproduce the S-matrix of string theory \cite{BSLV}. The details of this dimensional reduction will be discussed later.
We need however to keep in mind that $Y^D=\theta$ has to be treated on different footing from $Y^\mu ,~~\mu =0,..,D-1$.
Then the ERG is (suppressing $\tau$) :
 \[
\int du ~ {\p L[X(u)]\over \p \tau} =
\]
\be	\label{ERG}
 \int dz~\int dz'~\hf~\dot G^{\mu \nu}(z,z')\Bigg( \int du~{\delta^2L[X(u)] \over \delta X^\nu(z')\delta X^\mu(z)}+
\int du~\int dv~ {\delta L[X(u)]\over \delta X^\mu(z)}{\delta L[X(v)]\over \delta X^\nu(z')}\Bigg)
\ee

 Here $\dot G^{\mu\nu} \equiv {\p G^{\mu\nu}\over \p \tau}$.
 
 \subsection{Functional Derivatives and Loop Variables}
 
When we use this equation in the loop variable formalism we have to generalize the equation. This is done as follows:
In the loop variable formalism for open strings we have an infinite number of "time coordinates" $\xn ~, n=1,2,...$.
\be   \label{LV}
\lpp
\ee
with
\be	\label{k}
k(t) = \ko + {\ki \over t} + {\kt \over t^2} +....+{\kn\over t^n}+...
\ee
and 
\be	\label{al}
\al(t)=e^{\sum x_n t^{-n}}\equiv 1+ {\al_1\over t}+...+{\aln\over t^n}+...
\ee
$\aln$ satisfy:
${\p \aln\over \p x_p}= \al_{n-p}$.

In open string theory we had introduced $Y(z,\xn)$ (we suppress $\mu$ index for convenience).
\be	\label{Y}
Y\equiv X(z) + \al _1 \p_zX(z) + \al _2  \p_z^2 X(z) + {\al _3 \p_z^3X(z)\over 2!}+...+{\aln \p_z^n X(z)\over (n-1)!}+...
\ee

with $Y_n= {\p Y\over \p \xn}$. 
Thus 
\be
\lpp = \gvk
\ee
Furthermore in I we had introduced  $Y_{n_1,n_2}= {\pp Y\over \p _{x_{n_1}}\p_{x_{n_2}}}$ and so on. 
 For closed strings we have in addition, $ \bar \aln ,  \bar x_n , ~n=1,2,...$. Thus we will let 
$z$ stand for the full set $\{ z, \xn , \zb , \bar x _n\}$. 
Also $Y(z,\xn)$ will be extended to $Y(z,\zb, \xn, \xnb)$ and additionally  $Y_{\bar n} = {\p Y \over \p \xnb}$ and $Y_{\bar n_1,\bar n_2}= {\pp Y\over \p _{\bar x_{n_1}}\p_{\bar x_{n_2}}}$, and mixed derivatives, $Y_{n;\mb}= {\pp Y \over \p _{\xn} \p _{\xmb}}$ and also higher mixed derivatives, $Y_{n_1,n_2;\mb _1,\mb _2}$ etc. The closed string loop variable is described below.

Thus in the (\ref{ERG}) we will read $Y(z)$ for $X(z)$, with the meaning of $z$ generalized as above.
 We can define a cutoff Green function $\lan Y(z) Y(z')\ran = G(z,z';\tau)$ (where also by $z$ we mean the full set $\{z,\zb,\xn,\xnb\}$). Finally we can define the delta function 
\[
\delta (z-z')\equiv \delta^2(z-z') \prod _{n=1,2...} \delta (\xn - x_n') \delta (\xnb - \bar x_n')\] and
\[dz \equiv dz ~d\zb~\prod _{n=1,2...}d\xn~d \xnb \]

Then we can define the functional derivatives in the usual way:
\[ {\delta \over \delta X(z)} X(u) = \delta (z-u) \] but now with the generalized meaning for the fields, coordinates and delta function.
Apply this to functional  $\int du~L[Y(u),Y_{n,\mb}(u)]$: ($\xn$ will be associated with $u$, $\xn'$ with $z'$ and $\xn ''$ with $z''$)
\[
{\delta \over \delta Y(z')} \int du~ L[Y(u),Y_{n;\mb}(u)]=\]
\[
\int du~\Big\{ {\p L [Y(u),Y_{n;\mb}(u)]\over \p Y(u)} \delta (u-z')+\]
\[
\sum_ {n=1,2,...} {\p L [Y(u),Y_{n;\mb}(u)]\over \p Y_n(u)} \p_{\xn}\delta (u-z')+
\sum_ {n_1,n_2=1,2,...} {\p L [Y(u),Y_{n;\mb}(u)]\over \p Y_{n_1,n_2}(u)} \p_{x_{n_1}}\p _{x_{n_2}}\delta (u-z')\]
\[
+\sum_ {\mb =1,2,...}
 {\p L [Y(u),Y_{n;\mb}(u)] \over \p Y_{\nb} (u) }
 \p_{\bar x _m}\delta (u-z') 
  +\sum_ {\mb_1,\mb _2=1,2,...}
 {\p L [Y(u),Y_{n;\mb}(u)] \over \p Y_{\mb_1,\mb_2} (u) }
 \p _{\bar x_{m_1}}\p_{\bar x_{m_2}} \delta (u-z')
+\]\be    \label{FD}
\sum_ {n,\mb=1,2,...}
 {\p L [Y(u),Y_{n,\mb}(u)] \over \p Y_{n,\mb} (u) }
 \p _{ x_{n}}\p_{\bar x_{m}} \delta (u-z') +...\Big\}\ee
 We have refrained from writing a completely general expression so as not to clutter the equation, but
 the pattern should be clear to the reader. 

\subsection{Closed Strings and Mixed Derivative Vertex Operators}
 
 We now apply this to the closed string action. We need to specify the Lagrangian and then apply the ERG (\ref{ERG}).  In the loop variable formalism for closed strings normally we would start with 
 \be \label{CLVnaive}
e^{i\ko . X (z) + \oint_c dt~ k(t) \al(t) \p_z X(z+t) + \oint_c d\bar t~ \bar k(\bar t) \bar \al(\bar t) \p_\zb X(\zb+\bar t)}
\ee

In earlier papers on loop variables for closed strings \cite{BSCS} we used the notation $\bar k_n$. We will use $k_\nb$ in this paper.  This is convenient here, because we are soon going to introduce loop variables with mixed indices.
Thus (\ref{CLVnaive}) gives:
 \be   \label{naive}
 \int dz~e^{i(\ko .Y + \sum_{n,\nb=1,2,...} (\kn . \yn +  k_\nb . Y_{\nb}) )}  = \int dz~\Big( \e (1+ i \kn . \yn + i k_{\nb}.Y_{\nb} - k_{n\mu} k_{\mb \nu} \yn ^\mu Y_{{\mb}}^\nu +...\Big)
 \ee 
 If in the above set we restrict ourselves to the vertex operators satisfying $L_0=\bar L_0$, that would give the complete
 closed string Lagrangian level by level in the old covariant (Polyakov) formalism. In the BRST formalism there are states involving the ghost oscillators. Corresponding to these, in the loop variable formalism there are the generalized loop momenta with Lorentz index in the extra dimension. These were called $q_n, \bar q_n$ in the earlier papers, and will be called $q_n,q_\nb$ here. It turns out that we also need vertex operators involving mixed derivatives. In \cite{BSCS} where the lowest level (graviton)
 free equation was derived it was seen that there were essential contributions from terms involving ${\pp \s\over \p z\p \zb}$. The Liouville mode does not in general factorize into holomorphic and anti holomorphic parts. This is also reflected in the observation made in the introduction that the regulated Green function (eg $ln~(z\zb + a^2)$) does not split into holomorphic and anti holomorphic part, since dependence on the Liouville mode arises when we regulate the theory. Thus as long as we have a finite cutoff it is not correct to impose $\p \bar \p X=0$ and therefore in the ERG we have to introduce mixed derivative vertex operators for consistency. 
 
 Indeed when one applies the techniques used in I and II for obtaining gauge invariant ERG in open strings, it becomes clear that we need these extra operators.  We recapitulate the basic idea as applied now to closed strings: The gauge variation of the Lagrangian at level $N = \{n,\mb\}$ has to be derivatives of  lower level terms in the Lagrangian. Thus if $L_N$ denotes the
 Lagrangian at level $N$ its gauge variation has to be of the form:
 \be \label{LN} \delta L_N =  \sum _{n,\nb =1,2,...}\lambda _n {\p L_{N-n}\over \p \xn} + \lambda _\nb {\p L_{N-\nb}\over \p \xnb}
 \ee
 In the case of open strings this had the consequence that we had to introduce separately $Y_{n,m,...}= {\p^{n+m+...}Y\over \p \xn \p \xm ...}$ although in the original loop variable formalism $Y_{n,m}=Y_{n+m}$. 
 We refer the reader to II for the full details of this construction at all levels, involving vertex operators of the form $K_{n,m...\mu} Y_{n,m,..}^\mu $. The Lagrangian at any level involves products of these. They can be obtained by
 expanding the generalized loop variable vertex operator:
 \[ e^{i\Big( \kom Y^\mu + \sum _{n,m,..}K_{n,m,..\mu} Y_{n,m,...}^\mu\Big)}\]

 Thus we had at level two \footnote{Level 0 is the tachyon which has no gauge transformation properties, so we ignore it in this paper.}
 \[L_2=(iK_{2\mu} \ytm + iK_{11\mu} Y_{11}^\mu - \hf \kim \yim \kin \yin) \e\] with $K_{2\mu} = y_2 \kom,~~ K_{11\mu} = \ktm - y_2 \kom$. These were defined with simpler gauge transformation properties: $\delta K_2 = \lambda_2 \ko, ~~\delta K_{11}= \li \ki$. (This is summarized in the next subsection below.)
   The gauge transformation of this level two Lagrangian ($L_2$) is:
  \[\delta L_2= \li {\p \over \p x_1}\underbrace{(i\kim {\p \yim\over \p x_1 }\e )}_{L_1}+ \lt {\p \over \p x_2} \underbrace{( \e )}_{L_0}\]  which is of the form in (\ref{LN}) specialized to open strings where we only have $\la _n$. 
 
 In the case of closed strings, we need not only $Y_{n,m..}$ but also $Y_{n,m..;\nb,\mb,..}$.
 This is clear from (\ref{LN}), where clearly there are mixed derivatives. Thus (\ref{naive}) is generalized to
 \be
 \int dz~e^{i\Big( \kom Y^\mu + \sum _{\{n,m,..\nb,\mb,..\}=1,2...}K_{n,m,..;\nb,\mb,...\mu} Y_{n,m,...;\nb,\mb,...}^\mu\Big)}
 \ee
 
 Now in II we had explicit expressions for $K_{n,m,...}$. In the notation of this paper $K_{n,m,...}=K_{n,m,..;0}$. The complex conjugate $K_{0;\nb,\mb,...}$ is obviously given by the same expression, complex conjugated.  Let us proceed to obtain expressions for the remaining $K$'s. 
\subsubsection{Loop Variables for closed strings}

We have seen the need for mixed derivatives. Thus the loop variable defined in \cite{BSCS} which was a simple generalization of the open string loop variable is not sufficient. Motivated by this we will generalize our loop variable. 
\[
Exp ~\Big(i\Big(\ko . X (z) + \oint_c dt~ k(t) \al(t) \p_z X(z+t) + \oint_c d\bar t~ \bar k(\bar t) \bar \al(\bar t) \p_\zb X(\zb+\bar t)+ 
\]
\be	\label{LVC}
+\oint_c dt\oint_c d\bar t~K(t,\bar t) \al(t)
\bar \al (\bar t) \p_z \p_\zb X(z+t,\zb+\tb)\Big)\Big)
\ee
Expansion for $k(t),\al(t)$ are as given earlier and $\bar k(\tb) ,\bar \al(\tb)$ are anti-holomorphic versions of the same.
The first three terms in the exponent are the terms given in (\ref{naive}). The fourth term involves $K(t,\tb)$ defined below:
\be	\label{K}
K(t,\tb)\equiv K_{0;0}+\sum _{\mb=1}^\infty K_{0;\mb}\tb^{-\bar m} + \sum _{n=1}^\infty K_{n;0}t^{-n} + \sum _{n=1,\mb=1}^\infty K_{n;\mb}t^{-n}\tb^{-\mb}
\ee
Expanding $X(z+t,\zb+\tb)$ gives
\be
\p_z\p_\zb X(z+t,\zb+\tb)=\p_z\p_\zb X+ t \p_z^2\p_\zb X+ \tb \p_z \p_\zb^2 X + t\tb \p_z^2\p_\zb^2X+ t^2 {\p_z^3\p_\zb X\over 2!}+t^2 {\p_z\p_\zb^3X\over 2!}+...
\ee
Plugging all this in (\ref{LVC}) gives:
\[ \ko \Big(X+ \al_1 \p_zX + \al_2 \p_z^2 X +{\al_3\p_z^3X\over 2!}+...+\bar \al_1 \p_zX + \bar \al_2 \p_\zb^2X+...+{\al_n\bar \al_m\p_z^n\p_\zb^mX\over (n-1)!(m-1)!}+..\Big)\]
\[+\underbrace{K_{1;0}}_{=k_1} \Big(  \p_zX + \al_1 \p_z^2X + {\al_2\p_z^3X\over 2!}+..+\bar \al_1\p_z\p_\zb X+
\bar \al_2\p_z\p_\zb^2X +\]\[...+\al_1\bar \al_1 \p_z^2\p_\zb X+{\al_2 \bar \al_1 \p_z^3\p_\zb X\over 2!}+...+\al_1\bar \al_2\p_z^2\p_\zb^2X+...+{\al_n \bar \al_m \p_z^{n+1}\p_\zb^m X \over n!(m-1)!}+...\Big)+
\]
\[...+K_{n;\mb}\Big({\p_z^n\p_\zb^mX\over (n-1)!(m-1)!}+ {\al_1\p_z^{n+1}\p_\zb^mX\over (n)!(m-1)!}+...+{\al_p\bar \al_q\p_z^{n+p}\p_\zb^{m+q}X\over (n+p-1)!(m+q-1)!}+...\Big)\]

If we define the coefficient of $\ko$ to be $Y$, (\ref{LVC}) can be compactly written as
\be  \label{GVC}
Exp\Big(i\Big( \ko.Y + K_{1;0}.{\p Y\over \p x_1}+K_{0;\bar 1}.{\p Y\over \p \bar x_1}+K_{1;\bar 1}.{\pp Y\over \p x_1\p \bar x_1}+...
+K_{n;\mb}.{\p^2Y\over \p \xn \p\bar x_m}+...\Big)\Big)
\ee
As for open strings${\p^4 Y\over \p x_{n_1} \p x_{n_2}\p \xb _{m_1}\p \xb_{m_2}}={\pp Y\over \p x_{n_1+n_2}\p x_{m_1+m_2}}$.
Again, just as for the open string we will nevertheless introduce separately, $K_{n_1,n_2,...;\bar m_1,\bar m_2,...}$ as the coefficient
of ${\p\over \p x_{n_1}}{\p\over \p x_{n_2}}...{\p\over \p \bar x_{m_1}}{\p \over \p \bar x_{m_2}} ...Y$.
Expressions for $K_{[n]_i;[\mb]_j}$, where $[n]_i$ denotes a particular partition of $n$, (i.e. $\{n_1,n_2,...\}:n_1+n_2 +...=n$),  will be given in the next subsection. 

 \subsection{Constructing $K$'s for closed strings}
 \subsubsection{$K$'s for open strings - Recapitulation}
 
 Let us first recollect the construction of $K_{n,m,...}$ in II. 
 If $[n]_i$ defines a particular partition of the level $N$, at which we are working, then 
\be	\label{Genrule}
\delta K_{[n]_i\mu} = \sum_{m\in [n]_i}\la_mK_{[n]_i/m~\mu}
 \ee
where $[n]_i/m$ denotes the partition with $m$ removed, and the sum is over {\em distinct $m$'s}. Thus for eg.
\[ \delta K_{m,n}= \la _m K_n + \la _n K_m\] and similarly
\[ \delta K_{m,m}=\la _m K_m\]

As explained in I and II, the construction of the $K_n$'s uses crucially the loop variable momentum in the internal direction, $q(t)$ defined as
\[ q(t) = q_0 + {q_1\over t} + ....+{q_n\over t^n}+...\] and also the fact that for levels grater than one $q_0 >0$. One then defines $\bar q_n, y_n$ as follows:

Define 
\be \label{qbar}
\bar q(t) \equiv {1\over q_0}q(t) = 1+ {\bar q_1\over t}+ {\bar q_2\over t^2}+...+{\bar q_n\over t^n}+... 
\ee
Note that this definition makes sense only if $q_0\neq0$. For massless states this variable is not defined. Thus, for instance, 
$\bar q_1$ by itself is not defined for open strings. Fortunately this is not needed (for open strings) - only the higher levels $\bar q_1^2$ etc, are needed. 
\be 	\label{y}
= e^{\sum _n y^n t^{-n}}= 1+ {y_1\over t} + {y_2+{y_1^2\over 2}\over t^2}+{y_3+y_1y_2+{y_1^3\over 6}\over t^3}+....
\ee
If we solve for $y_n$ in terms of $q_m$ we get
\[ \bar q_1 = y_1;~~~\bar q_2= y_2+{y_1^2\over 2} \implies y_2= \bar q_2- {\bar q_1^2\over 2};\] Similarly \[y_3=\bar q_3 - \bar q_2\bar q_1 + {\bar q_1^3\over 3}\] In general
$\sum_{n=0}^\infty {y_n \over t^n}= ln~ (\bar q(t))$. 

The gauge transformation of $q(t)$  (see I,II and \cite{BSLV}) is $q(t)\rightarrow q(t) \la (t)$, where 
\[\la (t) = 1+{\li\over t} +...+{\la _n\over t^n}+...=e^{\sum_n z^nt^{-n}}\] Thus the gauge transfromation on $y_n$ is
$y_n \rightarrow y_n+z_n$.
To linear order in $\la _m$ this becomes: \[ \delta y_n = \la _n \] 
Thus the $K$'s can be defined as follows:
 
 \[K_{n,m,p,...\mu} = y_n y_my_p...\kom :~~~~n,m..\geq 2,~~n\neq m \neq p...\]
 For repeated indices the rule is 
 \[ K_{\mu {\underbrace{mmm..}_{i ~times}},{n,p...}} = {y_m^i\over i!}y_n y_p...\kom,~~~m,n,p,..\geq2,~~m\neq n\neq p...\]

When some of the indices are equal to 1:
\[ K_{n,m,p,...\underbrace{11...1}_{i~ times}\mu} = y_n y_m y_p...K_{\mu_{\underbrace{11..1}_{i ~times}}},~~n\neq m\neq p...~~~n,m,p...\geq 2\]

Again with repeated indices:

\[ K_{\mu \underbrace{m,m,..}_{j ~times},p,...\underbrace{11...1}_{i~ times}} = {y_m^j\over j!} y_p...K_{\mu \underbrace{11..1}_{i ~times}},~~ m\neq p...~~~m,p...\geq 2\]

This recursively defines all the $K$'s provided we give a prescription for $K_{11...1\mu}$:
\[ K_{1\mu} = \kim ~~K_{11\mu} = \ktm - K_{2\mu}, ~~~K_{111\mu} = k_{3\mu} - K_{21\mu} - K_{3\mu}\]

The general rule proved in II is
\be
K_{\mu\underbrace {1....1}_{n}} = k_{n\mu} - \sum_{[n]_i\in [n]' }K_{[n]_i\mu}
\ee
where $[n]'$ indicates all the partitions of $n$ {\em except} $\underbrace{1...1}_{n}$.

\subsubsection{$K$'s for closed strings}

Finally for closed strings, we make the identification
\[ K_{\mu n,m...;0}=K_{\mu n,m,..}\] where the RHS are the $K$'s that we have just defined. Similarly
$K_{\mu 0;\nb,\mb...}$ is given by the same expressions with bars i.e. $\nb$ instead of $n$, $k_{\bar 1\mu}$ instead of $\kim$ etc.

Now we come to the mixed $K$'s:
\be K_{1;\bar 1 \mu} = \bar y_1 \kim + y_1 k_{\bar 1 \mu} -y_1 \bar y_1 \kom  = \bar y_1 K_{1;0\mu} + y_1 K_{0;\bar1\mu} - y_1 \bar y_1 \kom\ee
One can check that \[ \delta K_{1;\bar 1\mu}= \la _1 k_{\bar 1\mu} + \bar \la _1 \kim \]

Similarly
\[ K_{1,1;\bar 1\mu} = \bar y_1 (\ktm - y_2 \kom ) + {y_1^2\over 2} k_{\bar 1\mu} - {y_1^2\over 2} \bar y_1 \kom \]
This can be rewritten as 
\be K_{1,1;\bar 1\mu}=\bar y_1 K_{1,1;0\mu} +{y_1^2\over 2} K_{0;\bar1\mu} -{y_1^2\over 2} \bar y_1 \kom\ee
One can easily verify that
\[ \delta K_{1,1;\bar 1\mu}= \la _1 K_{1;\bar 1\mu} + \bar \la _1 K_{1,1;0\mu}\] as required.
The pattern is clear:
\be
K_{\underbrace{1,1,...1}_{n};\underbrace{\bar 1,\bar 1,...\bar 1}_{m}\mu} = {\bar y_1^m\over m!}K_{\underbrace{1,1,...,1}_{n};0\mu} + { y_1^n\over n!}K_{0;\underbrace{\bar 1,\bar1,...,\bar1}_{m}\mu} - {y_1^n\over n!}{\bar y_1^m\over m!}\kom
\ee

Let us check the variation:
\[\delta K_{\underbrace{1,1,...1}_{n};\underbrace{\bar 1,\bar 1,...\bar 1}_{m}\mu}= \bar \la _1\Big({\bar y_1^{m-1}\over (m-1)!}K_{\underbrace{1,1,...,1}_{n};0\mu} + { y_1^n\over n!}K_{0;\underbrace{\bar 1,\bar1,...,\bar1}_{m-1}\mu} -{y_1^n\over n!}{\bar y_1^{m-1}\over (m-1)!}\kom \Big) +\]\[ \la _1\Big({\bar y_1^m\over m!}K_{\underbrace{1,1,...,1}_{n-1};0\mu} +{ y_1^{n-1}\over (n-1)!}K_{0;\underbrace{\bar 1,\bar1,...,\bar1}_{m}\mu}-{y_1^{n-1}\over (n-1)!}{\bar y_1^m\over m!}\kom \Big)
\]
\[= \bar \la _1K_{\underbrace{1,1,...1}_{n};\underbrace{\bar 1,\bar 1,...\bar 1}_{m-1}\mu}+\la _1K_{\underbrace{1,1,...1}_{n-1};\underbrace{\bar 1,\bar 1,...\bar 1}_{m}\mu} \]

One can then check that 
\be K_{p_1,p_2,..,\underbrace{1,1,...1}_{n};\bar q_1,\bar q_2,..,\underbrace{\bar 1,\bar 1,...\bar 1}_{m}\mu} = y_{p_1}y_{p_2}... \bar y_{q_1}\bar y_{q_2}.. K_{\underbrace{1,1,...1}_{n};\underbrace{\bar 1,\bar 1,...\bar 1}_{m}\mu} \ee with \[~~p_1,p_2,...,q_1,q_2,; \geq 2,~~~p_1\neq p_2\neq...; \bar q_1 \neq \bar q_2 \neq ...
\]
has the right gauge transformation. If any of the $p$ are repeated $i$ times, then $y_p$ is replaced by $y_p^i\over i!$. Similarly for the $\bar y_q$.

This completes the construction of $K_{\mu [n];[\mb]}$.   The loop variables that are involved are the same as for the physical vertex operators of closed string theory, so no new degrees of freedom have been added. However,
in principle one could add to $K_{[n]_i;[\mb]_j \mu}$, new variables of the form $k_{[n]_i;[\mb]_j \mu}$ with transformation rule 
\be \label{knmb}
\delta k_{[n]_i;[\mb]_j \mu} = \la _p k_{[n]_i/p;[\mb]_j \mu} + \bar \la _p k_{[n]_i;[\mb]/\bar p \mu}
\ee 
where as earlier $[n]_i/p$ stands for the particular partition $[n]_i$ with the one $p$ removed. (If $[n]_i$ does not contain $p$, that term does not contribute to the gauge transformation, and can be set to zero.) In fact we will do precisely this later on, although the extra variable will be determined algebraically by the existing ones, so new degrees of freedom are still not being added.

\subsubsection{An interesting relation}

The $K$'s obey an interesting relation of the form:
\be	\label{intrel}
\tilde K_{n;\mb \mu} \equiv \sum _{i,j} K_{[n]_i;[\mb]_j \mu} = \bar q_n k_{\mb \mu} + \bar q_{\mb} k_{n\mu} - \bar q_n \bar q_{\mb} \kom
\ee
Here, as earlier $[n]_i$ denotes a particular partition of $n$ denoted by $i$ and $\bar q$ was defined in (\ref{qbar}). Thus for instance
\[
\tilde K_{2,\bar 1\mu} \equiv K_{2;\bar 1\mu} + K_{1,1;\bar 1\mu} = \bar q_2 k_{\bar 1\mu} + \bar q_{\bar 1} \ktm - \bar q_{\bar 1} \bar q_2\kom\] 

The gauge transformation of $\tilde K_{n;\mb \mu}$ under $\la _p$ is easily seen to be:
\be
\delta \tilde K_{n;\mb \mu} = \la _p \tilde K_{n-p;\mb \mu}
\ee
This can be reasoned as follows: The only partitions $[n]_i$ that contribute to the gauge transformation, are the ones that have at least one $p$. Take these partitions and remove one $p$. The remaining numbers are all possible ways of making $n-p$ - so we get all the partitions of $n-p$. The gauge transformation law then forces (\ref{intrel}) to be true.
This relation will be used in the construction of the free equations. 

For the free equations one has to keep
only single derivatives in the loop variable. Thus we write ${\p \over \p x_{n_1+n_2+..}}{\p\over \p x_{\mb_1+\mb_2+..}}Y$ for $\p_{n_1}\p_{n_2}... \p_{\mb_1} \p_{\mb_2}Y...$. Thus the coefficient of  $Y^\mu_{n_1+n_2+...;\mb_1+\mb_2+..}$ is $\tilde K_{n;\mb \mu}$

Of course one can still add some new variables $k_{[n]_i,[\mb]_j \mu}$ with the correct gauge transformation law (\ref{knmb}), as mentioned earlier and then this would contribute to $\tilde K_{n,\mb \mu}$ also. This is in fact done in Appendix B.
 
 \section{Gauge Invariant Equations and the Problem of the Massive Graviton}
 
 We now proceed to evaluate the ERG acting on the closed string world sheet action. The free part involves second derivatives and is evaluated in the Appendix. As in the case of open strings, the interacting part is in the form of products of two gauge invariant field strengths. These are evaluated below and the gauge invariance is manifest. One then has to evaluate the OPE of these in the standard fashion. We do not work this out since the details are not really important at this point.

 \subsection{Level 2 (1;1)}

We start with level 2. The interaction Lagrangian $L$ at level 2 is best obtained by starting with the generalized loop variable, which we denote by $\cal L$. 
\be
{\cal L} = e^{i\Big( \ko.Y + K_{1;0}Y_{1;0}+K_{0;\bar 1}Y_{0;\bar 1} + K_{1;\bar 1}Y_{1;\bar 1} + K_{2;0}Y_{2;0}+K_{0;\bar 2}Y_{0;\bar 2} + K_{1,1;0}Y_{1,1;0}+K_{0;\bar1,\bar1}Y_{0;\bar 1,\bar 1}+....\Big)}
\ee 

Let us evaluate the functional derivative (\ref{FD}) on $\cal L$. Acting once it gives the gauge invariant field strength. Since we need $L_0=\bar L_0$ for the Lagrangian, we can
act on $\cal L$ and extract terms proportional to $k_{1\mu} k_{\bar 1\nu}$ or $K_{1;\bar 1\mu}$, and satisfying
$L_0=\bar L_0$. This means that the vertex operator has to be $\yim Y_{\bar 1}^\nu$ or $Y_{1; \bar 1}^\nu$.
The latter will drop out of all integrated correlations functions in the continuum limit so we can drop those terms.
\subsubsection{Field Strength}
We get 
\[
 \int du~ {\delta \over \delta Y^\mu(z')} {\cal L}(u)=\int du~\Big\{ {\p {\cal L} [Y(u),Y_{n,\mb}(u)]\over \p Y^\mu(u)} \delta (u-z')+{\p {\cal L} [Y(u),Y_{n;\mb}(u)]\over \p Y_1^\mu(u)} \p_{x_1}\delta (u-z')
\]\[+ {\p {\cal L}[Y(u),Y_{n;\mb}(u)] \over \p Y_{\bar 1} ^\mu(u) }
 \p_{\bar x _1}\delta (u-z') + {\p {\cal L}[Y(u),Y_{n,\mb}(u)] \over \p Y_{1,\bar 1}^\mu (u) }
 \p _{ x_{1}}\p_{\bar x_{1}} \delta (u-z') \Big\}
 \]
 \[
 = \int du~\Big\{ {\p {\cal L} [Y(u),Y_{n,\mb}(u)]\over \p Y^\mu(u)} \delta (u-z')-[\p_{x_1}{\p {\cal L} [Y(u),Y_{n;\mb}(u)]\over \p Y_1^\mu(u)}] \delta (u-z')
\]
\[- [\p_{\bar x _1} {\p {\cal L}[Y(u),Y_{n;\mb}(u)] \over \p Y_{\bar 1} ^\mu(u) }]
\delta (u-z') +  [\p _{ x_{1}}\p_{\bar x_{1}}{\p {\cal L}[Y(u),Y_{n,\mb}(u)] \over \p Y_{1,\bar 1}^\mu (u) }]
 \delta (u-z') +
\]
\[
  [\pp _{ x_{1}}{\p {\cal L}[Y(u),Y_{n,\mb}(u)] \over \p Y_{1,1;0}^\mu (u) }]\delta(u-z')+ [\pp _{ \bar x_{1}}{\p {\cal L}[Y(u),Y_{n,\mb}(u)] \over \p Y_{0;\bar1,\bar 1}^\mu (u) }]\delta(u-z')
  \]
  \[- [\pp _{ x_{1}}\p_{\bar x_ 1}{\p {\cal L}[Y(u),Y_{n,\mb}(u)] \over \p Y_{1,1;\bar 1}^\mu (u) }]\delta(u-z')- [\pp _{ \bar x_{1}}\p_{x_1}{\p {\cal L}[Y(u),Y_{n,\mb}(u)] \over \p Y_{1;\bar1,\bar 1}^\mu (u) }]\delta(u-z')
   \Big\}
 \]
 \be  \label{FnlDer}
 +
  [\pp _{ \bar x_{1}}\pp_{x_1}{\p {\cal L}[Y(u),Y_{n,\mb}(u)] \over \p Y_{1,1;\bar1,\bar 1}^\mu (u) }]\delta(u-z')
\ee
\[   
=
 \Big\{i\kom{\cal L }(z') -i K_{1;0\mu} \p_{x'_1} {\cal L}(z')-i K_{0;\bar 1\mu} \p_{\bar x' _1}  {\cal L}+iK_{1;\bar 1\mu} \p _{ x'_{1}}\p_{\bar x'_{1}} {\cal L}(z')+
 \]
 \be \label{FldStr}
  iK_{1,1;0\mu} \pp_{x_1}{\cal L}+ iK_{0;\bar 1,\bar 1}\pp_{\bar x_1}{\cal L}-iK_{1,1;\bar 1}^\mu \pp_{x_1}\p_{\bar x_1}{\cal L}-iK_{1;\bar 1,\bar 1}\p_{x_1}\pp_{\bar x_1}{\cal L}+iK_{1,1;\bar 1,\bar 1\mu} \pp_{x_1}\pp_{\bar x_1}{\cal L}\Big\}
 \ee
We have kept only terms that contribute to level $(1;\bar 1)$ and $(1,1;\bar 1,\bar 1)$. 
 From the structure of ${\cal L}$ we can see that 
 \be \label{Lvar}
 \delta {\cal L} = \sum_{n,\nb =1,2,...} (\la _n {\p \over \p \xn}{\cal L}+\la _{\bar n} {\p \over \p \xnb}{\cal L})
\ee
Using (\ref{Lvar})  we can easily check that (\ref{FldStr}) is invariant under $\la_1,\la _{\bar 1}$ variations, and at level 2, is the gauge invariant field strength for closed strings. We write it explicitly below:

\[
-i\kom ( K_{1;0}.Y_{1;0})(K_{0;\bar 1}.Y_{0;\bar 1})\e -\kom K_{1;\bar 1}. Y_{1;\bar 1}\e\]
\[
i K_{1;0\mu}( \ko. Y_{1;0})( K_{0;\bar1}.Y_{0;\bar 1}) \e +K_{1;0\mu} K_{0;\bar1}Y_{1,\bar 1}\e\]
\[
iK_{0;\bar 1\mu} (K_{1;0}.Y_{1;0})(\ko . Y_{0;\bar1})\e +K_{0;\bar 1\mu} K_{1;0}Y_{1;\bar 1}\e
\]
\be
-i K_{1;\bar 1\mu} (\ko Y_{0;\bar1})(\ko.Y_{1;0})\e - K_{1;\bar 1\mu} \ko.Y_{1;\bar1}\e
\ee

At level 2 the physical fields are the graviton, antisymmetric tensor and dilaton. Since $K_{1;\bar 1}$ involves $\bar q_1$, this field
strength is well defined only if the graviton and dilaton are massive and $q_0\neq0$. Thus as things stand, this cannot
describe the usual closed string states which are massless at this level. We will describe the resolution of this problem later.  

Let us write this equation in terms of space time fields and analyze the gauge transformations:
Define \[
\lan \hf k_{1(\mu} k_{\bar 1\nu)}\ran = h_{\mu \nu}~~~;
\] 
 \be
\lan \hf k_{1[\mu} k_{\bar 1\nu]}\ran = B_{\mu \nu}
\ee
 Let
us also define \[
\lan \hf(\la _1 k_{\bar 1\mu} + \bar \la _1 k_{1\nu})\ran = \eps _\mu ~~;
\] 
 \be
\lan \hf(\la _1 k_{\bar 1\mu} - \bar \la _1 k_{1\nu})\ran = \Lambda _\mu
\ee Then the gauge transformation laws are 
\[	
\delta_G h_{\mu \nu} = \p_{(\mu}\eps_{\nu)}  ~~~;
\] 
 \be	\label{gauge}
\delta_G B_{\mu \nu}=
 \p_{[\mu}\Lambda_{\nu]}
 \ee
  which are the expected forms for the linearized transformation for the metric perturbation and antisymmetric tensor associated with coordinate transformations. However, the non linear part, which should be a tensorial "rotation",  we don't see here. This is a problem if we are to identify these gauge transformations with general coordinate transformations. There is also another problem: 
 
  The coefficient of $Y_1^\mu Y_{\bar 1}^\nu$ can be seen to be
 \be	\label{fldstrlevel2}
 -\kor k_{1\mu} k_{\bar 1\nu} + k_{1\rho} \kom k_{\bar 1\nu} + k_{\bar 1\rho} k_{1\mu} \kon - K_{1;\bar 1\rho} \kom \kon 
 \ee 
In terms of space time fields this is 
\[
G_{\rho\mu\nu}\equiv\Big( -\p _\rho (h_{\mu \nu} + B_{\mu \nu})+ \p _\mu (h_{\rho \nu}+B_{\rho \nu})+ \p _\nu (h_{\mu \rho}+ B_{\mu \rho})\Big)- \p_\mu \p _\nu S_\rho
\] 
\be   \label{spin2}
=\Gamma_{\rho \mu \nu} + H_{\rho \mu \nu} - \p_\mu \p _\nu S_\rho
\ee
$H= dB$ is the gauge invariant 3-form field strength for $B$ and $\Gamma$ is the Christoffel connection for gravity. $\Gamma _{\rho \mu \nu} \rightarrow \Gamma _{\rho \mu \nu} + 2\p_\mu \p _\nu \eps _\rho$ is the linearized gauge transformation for the Christoffel connection. Since $S_\rho\rightarrow S_\rho + \lan \la _1 k_{\bar 1\rho} + \bar \la _1 \kim \ran = S_\rho + 2 \eps _\rho$ our "gauge invariant field strength" is indeed gauge invariant. This construction thus requires an auxiliary field $S_\mu$ that transforms by an inhomogeneous term (a shift). One could use this shift, $\eps$ gauge transformation, to gauge away this field ($S_\mu$). This would use up the gauge transformation and result in extra polarization components for $h_{\mu \nu}$. Thus in effect the graviton would be massive. This
is consistent with the observation made above that the construction of such an $S_\mu$ field is only possible if  it is massive and $q_0\neq 0$. Thus as it stands this theory cannot describe gravity. This is the second problem. It will turn out that these problems are interrelated.
This analysis also identifies the form of the gauge transformation of $\p_\mu \p_\nu S_\rho$ as being that of the Christoffel connection, a fact that will be extremely pertinent in the resolution of the two problems of the massive graviton and the Abelian gauge transformation.

\subsubsection{Free Equation}

The second functional derivative $\int dz'~\int dz''~\hf~\dot G(z',z'') \int du~{\delta^2L[X(u)] \over \delta X(z'')\delta X(z')}$ gives the free equation.

We need to evaluate
\[
\eta^{\mu\nu}{\delta \over \delta X^\nu(z')}
\int du~\Big\{ {\p {\cal L} [Y(u),Y_{n,\mb}(u)]\over \p Y^\mu(u)} \delta (u-z'')-\p_{x_1}{\p {\cal L} [Y(u),Y_{n;\mb}(u)]\over \p Y_{1;0}^\mu(u)} \delta (u-z'')
\]
\be - \p_{\bar x _1} {\p {\cal L}[Y(u),Y_{n;\mb}(u)] \over \p Y_{0;\bar 1} ^\mu(u) }
\delta (u-z'') +  \p _{ x_{1}}\p_{\bar x_{1}}{\p {\cal L}[Y(u),Y_{n,\mb}(u)] \over \p Y_{1,\bar 1}^\mu (u) }
 \delta (u-z'') \Big\}
\ee
\[
=\eta^{\mu\nu}\int du~[{\p\over \p Y^\nu(u)} +{\p\over \p x_1}\delta(u-z'){\p \over \p Y^\nu_{1;0}(u)}+{\p\over \p \bar x_1}\delta(u-z'){\p \over \p Y^\nu_{0;\bar1}(u)}+{\p\over \p x_2}\delta(u-z'){\p \over \p Y^\nu_{2;0}(u)}+\]\[{\p\over \p \bar x_2}\delta(u-z'){\p \over \p Y^\nu_{0;\bar 2}(u)}+
 {\pp\over \p x_1\p \bar x_1}\delta(u-z'){\p \over \p Y^\nu_{1;\bar 1}(u)}+...]\]
 \[
 \Big\{ {\p {\cal L} [Y(u),Y_{n,\mb}(u)]\over \p Y^\mu(u)} \delta (u-z'')-\p_{x_1}{\p {\cal L} [Y(u),Y_{n;\mb}(u)]\over \p Y_1^\mu(u)} \delta (u-z'')
\]
\be - \p_{\bar x _1} {\p {\cal L}[Y(u),Y_{n;\mb}(u)] \over \p Y_{\bar 1} ^\mu(u) }
\delta (u-z'') +  \p _{ x_{1}}\p_{\bar x_{1}}{\p {\cal L}[Y(u),Y_{n,\mb}(u)] \over \p Y_{1,\bar 1}^\mu (u) }
 \delta (u-z'') \Big\}
\ee
This is evaluated in the Appendix. The result is
\be
[-\ko^2 \kim k_{\bar 1\nu} + \ko .\ki \kom k_{\bar 1\nu} + \ko. k_{\bar 1}\kon \kim - K_{1;\bar1}.\ko \kom \kon]Y_{1;0}^\mu Y_{0;\bar 1}^\nu
\ee
If we use the constraint $K_{1;\bar1}.\ko = k_1.k_{\bar1}$ (see Appendices) this equation is just 
\[
[-\ko^2 \kim k_{\bar 1\nu} + \ko .\ki \kom k_{\bar 1\nu} + \ko. k_{\bar 1}\kon \kim - k_1.k_{\bar1} \kom \kon]Y_{1;0}^\mu Y_{0;\bar 1}^\nu
\]
\be	\label{grav}
\p^\rho \Gamma_{\rho \mu \nu}-\p_\mu\p _\nu h^\rho_\rho= -\pp h_{\mu\nu} + \p_\mu \p^\rho h_{\rho \nu} + \p_\nu\p^\rho h_{\rho \nu} - \p_\mu \p_\nu h^\rho_\rho=0 
\ee
 which is gauge invariant. This is the usual linearized equation for the metric perturbation.

\subsection{Level 4 (2;2)}

\subsubsection{Interacting Equation}

Let us evaluate the field strength proportional to $Y_{1;0}^\mu Y_{1;0}^\nu Y_{0;\bar 1}^\rho Y_{0;\bar 1}^\sigma \e$.
We can extract it from (\ref{FldStr}):

The field strength is given by:
\[
i\kom {(iK_{1;0}.Y_1)^2\over 2!}{(i K_{0;\bar 1}.Y_{\bar 1})^2\over 2!}
-iK_{1;0\mu} (i\ko.Y_1)(iK_{1;0} .Y_1){(iK_{0;\bar 1}.Y_{\bar 1})^2\over 2!} \]\[
-iK_{0;\bar 1\mu}(i\ko.Y_{\bar 1})(iK_
{0;\bar 1}.Y_{\bar 1}){(iK_{1;0}.Y_1)^2\over 2!}
+iK_{1;\bar 1\mu} (i\ko.Y_1)(i\ko.Y_{\bar 1} )(iK_{1;0}.Y_1)(iK_{0;\bar 1}.Y_{\bar 1})\]
\[+iK_{1, 1;0\mu} (i\ko.Y_1)^2{(iK_{0;\bar 1}.Y_{\bar 1})^2\over 2!}+iK_{0;\bar1, \bar 1\mu} (i\ko.Y_{\bar 1})^2{(iK_{1;0}.Y_{ 1})^2\over 2!}
\]
\[
-iK_{1,1;\bar 1\mu}(i\ko.Y_{\bar 1})(iK_{0;\bar 1}.Y_{\bar 1})(i\ko.Y_{ 1})^2
-iK_{1;\bar 1,\bar 1\mu}(i\ko.Y_{ 1})(i\ko.Y_{\bar 1})^2(iK_{ 1;0}.Y_{  1})
\]
\be	\label{int4}
+iK_{1,1;\bar 1,\bar 1\mu}(i\ko.Y_{\bar 1})^2(i\ko.Y_{ 1})^2
\ee
It is easily verified that it is gauge invariant.

Explicit expressions for the $K$'s is given in Section 2. $K_{1;0}= k_1$ and $K_{0;\bar 1}= k_{\bar 1}$.
\subsubsection{Free Equation}

The free equation is evaluated in the Appendix. The result is given below. We have assumed a metric for contraction of indices, this is discussed in the description of the ERG:
\[
-{1\over 4} \ko^2 (k_1 . Y_1)^2(k_{\bar 1}.Y_{\bar 1})^2 + \hf \ko.\ki (\ko . Y_1)(\ki .Y_1)(k_{\bar 1}.Y_{\bar 1})^2+\hf \ko.k_{\bar 1} (\ko . Y_{\bar 1})(k_{\bar 1} .Y_{\bar1})(k_{ 1}.Y_{1})^2
\]
\be 	
- {\ki .\ki\over 4} (\ko . Y_1)^2(k_{\bar 1}.Y_{\bar 1})^2 - {k_{\bar 1} .k_{\bar 1}\over 4} (\ko . Y_{\bar 1})^2(k_{ 1}.Y_{\ 1})^2 - \ki .k_{\bar 1} (\ko . Y_1)(\ko . Y_{\bar 1})(\ki .Y_1)(k_{\bar 1} .Y_{\bar1})
\ee
One can extract the coefficient of $Y_1^\mu Y_1^\nu Y_{\bar 1}^\rho Y_{\bar 1}^\sigma$ to get:
\[
-\ko^2\kim \kin k _{1\rho} k_{1\sigma} + \ko.\ki k_{0(\mu}k_{1\nu)}k_{\bar 1\rho} k_{\bar 1\sigma} + \ko .k_{\bar 1} k_{0(\rho}k_{1\sigma)}\kim \kin
\]
\be  \label{3243}
-\ki.\ki k_{\bar 1\rho} k_{\bar 1\sigma} \kom \kon - k_{\bar 1}.k_{\bar 1}\kim \kin k_{0\rho} k_{0\sigma} - \ki .\ki k_ {1(\mu} k_{0\nu)} k_{\bar 1(\rho} k_{0\sigma)}=0
\ee
Let us define $\lan \kim \kin k_{\bar 1\rho} k_{\bar 1\sigma}\ran= S_{\mu \nu \rho \sigma}$. This is a "spin 4" tensor symmetric in the first two and last two indices.

Defining
$ \p_{(\sigma|} \p_{(\nu}S_{ \mu)\la~|\rho)}^{~~~~\la} $ as the sum of four terms symmetrized in $\mu \nu$ and $\rho \sigma$ and
$\p^\la \p_{(\sigma} S_{|\mu \nu \la|\rho)}$ as the sum of two terms symmetrized in $\rho \sigma$, we can write this equation as
\[
-\pp S_{\mu \nu \rho\sigma} + \p^\la \p_{(\mu}S_{\nu) \la \rho \sigma} + \p^\la \p_{(\sigma} S_{|\mu \nu \la|\rho)}
\]
\be	\label{free4}
-\p_\mu\p_\nu S^\la_{ ~\rho \sigma \la} -\p_\rho \p_\sigma S_{\mu \nu \la}^ \la+\p_{(\sigma|} \p_{(\nu}S_{ \mu)\la~|\rho)}^{~~~~\la}
=0
\ee

The gauge transformation is $\kim \rightarrow \kim + \li \kom ;~~k_{\bar 1\mu} \rightarrow k_{\bar 1\mu}+\la_{\bar 1} \kom$.
Defining $\lan \la_1 \kim k_{\bar 1\rho} k_{\bar 1\sigma}\ran = \Lambda _{\mu \rho \sigma}$ and
$\lan \la _{\bar 1} \kim \kin k_{\bar 1\rho} \ran = \bar \Lambda _{\mu\nu \rho}$ the gauge transformation is

\be \label{gaugeS4}
\delta S_{\mu \nu \rho \sigma} = \p_{(\mu} \Lambda _{\nu) \rho \sigma} +\p_{(\rho|}\bar \Lambda _{\mu \nu | \sigma)}
\ee
The gauge parameter obeys a tracelessness constraint: the trace on any two indices is zero.
\subsection{Equation for Spin 4 Interacting with Two Gravitons}

We can write down a term in the full interacting equation corresponding to $Y_1^\mu Y_1^\nu Y_{\bar 1}^\rho Y_{\bar 1}^\sigma$ by combining (\ref{spin2}) and (\ref{free4}):
\[
\int dz~\dot G(z,z;\tau) (-\pp S_{\mu \nu \rho\sigma} + \p^\la \p_{(\mu}S_{\nu) \la \rho \sigma} + \p^\la \p_{(\sigma} S_{|\mu \nu \la|\rho)}
\]
\be	\label{Spin4}
-\p_\mu\p_\nu S_{\la ~\rho \sigma}^{~\la} -\p_\rho \p_\sigma S_{\mu \nu \la}^{~~~~\la}+\p_{(\sigma|} \p_{(\nu}S_{ \mu)\la~|\rho)}^{~~~~\la})
+ \int dz ~\int dz'~\dot G(z,z';\tau) G^\la _{~~\mu \rho} G_{\la \nu \sigma} +...=0
\ee
The three dots indicate other interactions. ($G_{\mu\nu\rho}$ is modified in the next section to (\ref{modspin2})). The terms described above correspond to a cubic interaction between two lowest level fields (either a graviton or antisymmetric tensor or dilaton) and a massive spin 4 field. Elucidating the full structure requires doing the dimensional reduction, which discussion we postpone. 

As mentioned earlier, the field strength $G_{\mu \nu \rho}$ will be modified in the next section. As things stand
the graviton described by this field strength is massive. But once $G_{\mu \nu \rho}$ is modified the above equation is correct.

\section{General Coordinate Transformation and Massless Graviton}

Let us restate our problems: One was that a gauge invariant EOM for the graviton can be written only when it is massive - the construction of the field $S_{\rho \mu \nu}$ in terms of loop variables required a non zero mass. The second problem is that the gauge transformations are Abelian and seem to have nothing to do with coordinate transformations. Let us focus on the second problem first.

\subsection{Combining Coordinate transformations and Gauge transformation}

The gauge transformation, which we call $\delta _G$, on the graviton has the form 
\[
\delta _G  h_{\mu \nu} = \tilde \eps _{(\mu ,\nu)}
\]
We should compare this with what we know from General Relativity (GR).
In GR the metric tensor obeys:
\[
g_{\mu \nu}(X) dX^\mu dX^\nu = g'_{\rho \sigma}(X')dX^{'\rho}dX^{'\sigma}
\]
We then find that (for {\em infinitesimal} $\eps$)
\be	\label{tensor}
\delta _{GCT} g_{\mu \nu}(X)\equiv g'_{\mu \nu}(X)-g_{\mu\nu}(X) =  \eps^\la g_{\mu\nu,\la}+  \eps ^\la_{~,\mu}g_{\la\nu}+ \eps^\la_{~,\nu}g_{\mu\la};~~~~\delta_{GCT}X\equiv X^{'\mu}-X^\mu = - \eps^\mu(X)
\ee
 The subscript "GCT" stands for General Coordinate Transformation. This is the standard tensor transformation law. We can now extract from this the transformation law for $h_{\mu \nu}$. Write $g_{\mu \nu}=\eta_{\mu \nu}+ h_{\mu \nu}$. Then (\ref{tensor}) becomes 
 \[
\delta _{GCT}g_{\mu \nu}=\delta _{GCT}h_{\mu \nu} = \eps ^\la h_{\mu \nu,\la}+ \eps ^\la_{~,\mu}(\eta_{\la \nu}+h_{\la \nu})+ \eps^\la_{~,\nu}(\eta_{\mu\la}+h_{\mu\la})
\]
\[=
\eps ^\la_{~,\mu}\eta_{\la \nu}+ \eps^\la_{~,\nu}\eta_{\mu\la}+ \eps ^\la h_{\mu \nu,\la}+\eps ^\la_{~,\mu}h_{\la \nu}+\eps^\la_{~,\nu}h_{\mu\la}
\]
\be 
\delta _{GCT}h_{\mu \nu}=
\eps_{(\mu,\nu)} + \eps ^\la h_{\mu \nu,\la}+\eps ^\la_{~,\mu}h_{\la \nu}+\eps^\la_{~,\nu}h_{\mu\la}
\ee
where we have defined $\eps _\mu \equiv \eta _{\mu \nu}\eps^\nu$. 

We see that $\delta _{GCT}$ acting on $h_{\mu \nu}$
 has two parts: an Abelian inhomogeneous term $\eps_{(\mu,\nu)}$ and a non-Abelian "rotation": $\eps ^\la h_{\mu \nu,\la}+\eps ^\la_{~,\mu}h_{\la \nu}+\eps^\la_{~,\nu}h_{\mu\la}$. This is analogous to Yang-Mills where there is an Abelian part to the gauge transformation $\delta A_\mu = \p _\mu \Lambda$ and a non-Abelian rotation $\delta A_\mu = \Lambda \times A_\mu$. The non-Abelian part is just a tensorial transformation.
 We will refer to this tensorial transformation as $\delta _T$, "T" stands for "tensor". Thus 
 \[
\delta _Th_{\mu \nu}\equiv \eps ^\la h_{\mu \nu,\la}+\eps ^\la_{~,\mu}h_{\la \nu}+\eps^\la_{~,\nu}h_{\mu\la};~~~~\delta_T X^\mu = -\eps ^\mu
\]
Thus acting on proper tensors, $\delta _T = \delta _{GCT}$. But on non tensorial fields such as $h_{\mu \nu}$ they are not the same.
Note however that \[
(\delta _G + \delta _T)h_{\mu \nu}=\tilde \eps_{(\mu,\nu)} + \eps ^\la h_{\mu \nu,\la}+\eps ^\la_{~,\mu}h_{\la \nu}+\eps^\la_{~,\nu}h_{\mu\la}
\]
Thus {\em if} we identify the gauge transformation parameter $\tilde \eps_\mu = \eps _\mu$, {\em then}, the combined transformation $(\delta _G+\delta _T) h_{\mu \nu} = \delta _{GCT}h_{\mu \nu}$. 

This suggests the following: It is easy to make EOM for proper tensor fields covariant under coordinate transformations. We can introduce a background metric and background covariant derivatives and write covariant EOM. But $h_{\mu\nu}$ is not quite a tensor. However we know that $\delta _G$ is already a symmetry of the EOM. So what is left is to implement $\delta _T$, which is a tensorial transformation, and therefore easy to implement. Then we should have general covariance for the $h_{\mu\nu}$ equation.

In order to implement $\delta _T$ we need to check if the Lagrangian is invariant. The interaction part $L_{int}$ is manifestly invariant: $h_{\mu \nu}\p_zX^\mu \p_\zb X^\nu$ is clearly invariant under $\delta_ T$. (This is also true for the massive higher spin fields with some modifications discussed later.) However the kinetic term $\eta _{\mu \nu}\p_zX^\mu \p_\zb X^\nu$ is not because $\eta _{\mu \nu}$ is a fixed matrix - and does not transform.

\be \label{kin}
\delta_T( \eta_{\mu \nu} \p_z X^\mu \p_\zb X^\nu) = -\eps_{(\mu,\nu)}\p_z X^\mu \p_\zb X^\nu
\ee

Non invariance of the kinetic term is very inconvenient because then the Green function is not covariant. Thus we make the kinetic term invariant and transfer this non invariance
to the interaction Lagrangian.

The kinetic term can be made invariant by the standard technique of
 introducing a background "reference" metric (and then covariantize the ERG) which we call $g_{\mu\nu}^R(X)$.  

For the purposes of this paper, we will keep the geometry flat, so that $g_{\mu \nu}^R$ is equivalent up to coordinate
transformation to $\eta_{\mu \nu}$. (Thus  for instance, for infinitesimal $\xi$,  $g_{\mu \nu}^R$ can be parametrized as 
$g_{\mu\nu}^R\equiv \eta_{\mu\nu}+h_{\mu\nu}^R=\eta_{\mu \nu} + \xi_{(\mu,\nu)}$,
where we take the new coordinates to be given by $X^{'\mu}=X^\mu - \xi^\mu(X)$.) 

Now let us add and subtract $h_{\mu \nu}^R(X)\p_z X^\mu \p_\zb X^\nu$ to the action:
\[
S =\int dz~ \underbrace{(\eta_{\mu\nu}+ h_{\mu \nu}^R(X))\p_z X^\mu \p_\zb X^\nu}_{L_{kinetic}}~+~
\underbrace{\overbrace{(h_{\mu \nu}(X)-h_{\mu \nu}^R(X))}^{\tilde h_{\mu\nu}}\p_z X^\mu \p_\zb X^\nu +....}_{L_{interactions}}
\]
The three dots stand for other vertex operators. Now let us perform the coordinate transformation $\delta _T X^\mu \equiv= -\eps^\mu$.
We will choose an action on $h^R_{\mu \nu}(X)$ such that the kinetic term is invariant.  Clearly we need to cancel (\ref{kin}) so
we need
\be
\delta _T (h^R_{\mu \nu}(X) \p_z X^\mu \p_\zb X^\nu)= \eps_{(\mu,\nu)}\p_z X^\mu \p_\zb X^\nu
\ee
This is more explicitly written as 
\be
\delta _T h^R_{\mu \nu}(X)=\eps^\la h^R_{\mu\nu,\la}+ \eps ^\la_{~,\mu}h^R_{\la\nu}+\eps^\la_{~,\nu}h^R_{\mu\la} + \eps_{(\mu,\nu)};~~~~\delta_TX = -\eps^\mu(X)
\ee
This gives the usual transformation of the background metric $g_{\mu \nu}^R$ under GCT. Thus we have $\delta _T L_{Kin}=0$. 
It is important to observe that although $g_{\mu \nu}^R$ is a background metric in the usual sense, $h_{\mu \nu}$ continues to defined as a fluctuation about $\eta _{\mu \nu}$, and not about $g_{\mu \nu}^R$. Thus $L_{int}$ involves $h_{\mu \nu} - h_{\mu \nu}^R$ now.
Another way of seeing this is that since we have added {\em and} subtracted $h_{\mu \nu}^R(X)\p_z X^\mu \p_\zb X^\nu$, we
really have not done anything physically different, so the physical interpretation of $h_{\mu \nu}$ remains the same.

Now $L_{int}$ is not invariant:
\[
\delta _T L_{int}=\delta _T [(h_{\mu \nu}(X)-h_{\mu \nu}^R(X))\p_z X^\mu \p_\zb X^\nu]=- \eps_{(\mu,\nu)}\p_z X^\mu \p_\zb X^\nu
\]

We can recover invariance if we {\em combine} $\delta _G$ with $\delta _T$ ,i.e. choose the gauge parameter $\tilde \eps_\mu$ to be
related to the coordinate transformation parameter $\eps ^\mu$: $\tilde \eps _\mu = \eta_{\mu \nu}\eps ^\nu \equiv \eps_\mu$\footnote{Note that $\eta_{\mu\nu}$ is being used in this equation and not $g_{\mu\nu}^R$}. Then 
using $\delta_G~h_{\mu\nu}(X)=\eps_{(\mu,\nu)}$ we get
\be
(\delta _T+\delta_G) L_{int}=(\delta _T+\delta_G) [(h_{\mu \nu}(X)-h_{\mu \nu}^R(X))\p_z X^\mu \p_\zb X^\nu]=0 
\ee
Thus $(\delta_G+\delta_T)L_{Kin}= 0=(\delta_G+\delta_T)L_{int}$. Since both terms are separately invariant, we can expect that the EOM obtained from the ERG will have this invariance manifest. Note that the combination $h_{\mu\nu}-h_{\mu\nu}^R\equiv \tilde h_{\mu\nu}$ transforms as a proper tensor under $\delta _T + \delta _G$. 

\subsection{Summary}
Let us summarize the above: Our starting point is
\[
Z[h_{\mu\nu},S_{\mu\nu\rho \sigma},...]= \int {\cal D}Xe^{\int dz \overbrace{(\eta _{\mu \nu}+ h_{\mu \nu})}^{g_{\mu\nu}}\p_zX^\mu \p_\zb X^\nu +S_{\mu\nu\rho \sigma}\p_zX^\mu \p_\zb X^\nu \p_zX^\rho \p_\zb X^\sigma+...}
\]

We would like to treat $h_{\mu \nu}$ as an interaction term. This is because we would like an exact RG equation that treats all the string modes in the same way. General covariance is then not manifest at any finite order. So we turn to a background field approach and introduce a non dynamical background $h_{\mu\nu}^R$ as follows:
\[=
\int {\cal D}Xe^{\int dz \overbrace{(\eta _{\mu \nu}+ h^R_{\mu \nu})}^{g_{\mu\nu}^R}\p_zX^\mu \p_\zb X^\nu +\overbrace{(h_{\mu \nu}-h^R_{\mu \nu})}^{\tilde h_{\mu\nu}}\p_zX^\mu \p_\zb X^\nu+
S_{\mu\nu\rho \sigma}\p_zX^\mu \p_\zb X^\nu \p_zX^\rho \p_\zb X^\sigma+...}
\]
\be
=
\int {\cal D}Xe^{\int dz  g^R_{\mu \nu}\p_zX^\mu \p_\zb X^\nu +\tilde h_{\mu\nu}\p_zX^\mu \p_\zb X^\nu+
S_{\mu\nu\rho \sigma}\p_zX^\mu \p_\zb X^\nu \p_zX^\rho \p_\zb X^\sigma+...}
\ee
Since the action does not depend on $h^R_{\mu\nu}$, the final answer should be independent of $h^R_{\mu \nu}(X)$. Thus formally 
\be	\label{hR}
{\delta Z \over \delta h_{\mu\nu}^R(k)}=0
\ee
Equivalently
\[
Z[g_{\mu\nu}^R,\tilde h_{\mu \nu}]=Z[g_{\mu\nu}^R+\tilde h_{\mu\nu}]=Z[g_{\mu\nu}]
\]
We have a {\em non dynamical} fixed background metric $g_{\mu \nu}^R = \eta_{\mu \nu}+h^R_{\mu \nu}$.
This will be treated as a background field and will be incorporated into the propagator. $g_{\mu\nu}^R,\tilde h$  are proper tensors under background GCT, which, in addition to transforming $h_{\mu\nu}$ (and the coordinate $X^\mu$), also includes transforming $h_{\mu\nu}^R$. (In fact $h$ and $h^R$ have similar transformations, which is why the difference $\tilde h_{\mu\nu}$ transforms as a proper tensor.) $g_{\mu\nu}^R$ is incorporated into the kinetic term and is treated non perturbatively, while $\tilde h_{\mu\nu}$
 will be treated perturbatively like any other 'matter' field  as an interaction term. 
 
 $Z$ is evaluated by expanding in a power series in $\tilde h$ and will be non polynomial in $\tilde h$. If we let $\tilde h_{\mu\nu}(X)= \int dk~ \tilde h_{\mu\nu}(k)e^{ik.X}$, we can think of $\tilde h_{\mu\nu}(k)$ as an infinite number of coupling constants.   There is a finite UV cutoff in place and the functional integral is well defined. 
Thus,  assuming a non zero  radius of convergence for the perturbation series, the formal statement (\ref{hR}) is expected to hold (on replacing $\tilde h= h-h^R$ in all the terms) after summing the series. We do not have a proof of this however.

Symmetry under GCT on the physical $h$ is not manifest when we treat $h$ perturbatively. However the background GCT symmetry, which also involves
$h^R$ is manifest order by order. This is because, as mentioned above, under background transformations $\tilde h$, and $g_{\mu\nu}^R$ are tensors and both interaction and kinetic terms are separately invariant. This is the gist of the earlier paragraphs of this section. 
 The EOM for $\tilde h$ can thus be made (background) covariant. 
This construction is very similar to what is done in the background field method for gauge theories \cite{Abbott}. There the field
$A_\mu$ is replaced by $\tilde Q_\mu + W_\mu$, where $W_\mu$ is an arbitrary background. The gauge transformation of the background field has the inhomogeneous term and $\tilde Q_\mu$ transforms homogeneously:
\[
\delta W_\mu = \p_\mu \Lambda + \Lambda \times W_\mu ;~~~~~\delta \tilde Q_\mu = \Lambda \times \tilde Q_\mu
\]
 The analog of $W+\tilde Q$ is $h_{\mu \nu}$. 
 
 Since the final answer for $Z$ cannot depend on $h^R$ (after the substitution $\tilde h=h-h^R$) because of (\ref{hR}), or equivalently, depends only on the sum $g^R+\tilde h$,  {\em background covariance of the equations also implies general covariance}.  
 
 In the $\beta$ function method, rather than $Z$ one calculates $\p Z\over \p (ln a)$. But the covariance arguments are the same.
When we include all other massive modes, one has to solve for all the massive modes in terms of $\tilde h$ first. This is equivalent to starting with the ERG that includes all the irrelevant coupling constants and obtaining the low energy $\beta$ function. (This is explained in \cite{W}). Assuming that the massive modes are all tensors, background covariance continues to hold order by order, and once again
if we invoke (\ref{hR}), this implies that the final low energy $\beta$ function for $h$ must be covariant. The role played by the background field is illustrated by a toy example in Appendix C.
 
 \subsection{Normal Coordinates}
 This subsection is a digression describing in some detail how normal coordinates are introduced. The main result is that one can obtain results that are manifestly background covariant. If the reader is willing to accept this, this section may be skipped.
 
 To obtain the background covariant equations one can use standard techniques of Riemann Normal Cooordinates \cite{AGFM,Pet,Eis}.
 However in our case since we start with a flat geometry, we can set all (background) curvature tensors to zero in their equations. This then becomes a rather trivial application of RNC. 

We begin with a point O with coordinates $x_0$ in some coordinate system, and another point P, with coordinates $x$. The geodesic
that connects the two points has a geometric meaning independent of the coordinate system. Similarly the tangent vector $\vec \xi$ (of unit length) to this geodesic at O is also a geometric object. One can derive from the geodesic equation \cite{Pet}:
\[
x^\mu = x_0^\mu + t \xi^\mu - {t^2\over 2!}\xi^\rho \xi ^\sigma \Gamma^\mu_{\rho \sigma} - {t^3\over 3!} \xi ^\rho \xi ^\sigma \xi ^\la
\Gamma ^\mu _{\rho \sigma \la} +.... 
\]
Here $t$ is the length of the geodesic from O to P and is also a geometric quantity. We can let $t\xi ^\mu = y^\mu$ which also has a geometric meaning independent of the coordinate system (it is a vector at O)  and write
\[
x^\mu (x_0,y)= x_0^\mu +y^\mu - {1\over 2!}y^\rho y^\sigma \Gamma^\mu_{\rho \sigma} - {1\over 3!} y ^\rho y^\sigma y ^\la
\Gamma ^\mu _{\rho \sigma \la} +.... 
\]
Finally we can introduce RNC, $Y^\mu=x_0^\mu+y^\mu$. The coordinate transformation from $x$ to $Y$ defines a matrix, which varies from point to point, 
$T^\mu_\rho (x)= {\p Y^\mu\over \p x^\rho}$ and can be used to transform tensors from one coordinate system to another.

In the RNC, $Y$, one can perform an ordinary Taylor expansion for any tensor $\bar W(Y)=\bar W(x_0+y)$ in powers of $y^\mu$ and obtain: 

\[
\bar W_{\al _1 ....\al _p}(Y) 
= \bar W_{\al _1 ....\al _p}(x_0) ~+~
\bar W_{\al _1 ....\al _p , \mu}(x_0)y^\mu ~+~
\]
\[
{1\over 2!}\{\bar W_{\al _1 ....\al _p ,\mu \nu}(x_0) 
~-~{1\over 3}
\sum _{k=1}^p \bar R^\beta _{~\mu \al _k \nu}(x_0) 
\bar W_{\al _1 ..\al _{k-1}\beta \al _{k+1}..\al _p}(x_0)\}
 y^\mu y^\nu
~+~
\]
\[   
{1\over 3!}\{\bar W_{\al _1 ....\al _p ,\mu \nu \rho}(x_0) - 
\sum _{k=1}^p \bar R^\beta _{~\mu \al _k \nu}(x_0) 
\bar W_{\al _1 ..\al _{k-1}\beta \al _{k+1}..\al _p, \rho}(x_0)
\]
\be \label{Taylor}
-
{1\over 2}\sum _{k=1}^p \bar R^\beta _{~\mu \al _k \nu ,\rho }(x_0) 
\bar W_{\al _1 ..\al _{k-1}\beta \al _{k+1}..\al _p}(x_0)\}y^\mu y^\nu y^\rho +...
\ee
where the derivatives are covariant derivatives. We have put bars over all the tensors to indicate that this expansion
requires a normal coordinate system. The LHS is a tensor at $Y=x_0+y$, whereas the RHS is a sum of tensors at $x_0$. Hence
this expansion makes sense only in this particular coordinate system. In the case of a scalar, the LHS is invariant: $\phi '(x')= \phi (x)$
and also every term on the RHS is individually a scalar and hence invariant. Thus if the expansion is true in one coordinate system it is true in any coordinate system.
For a tensor, under a coordinate change, the LHS will have to be multiplied by an appropriate number of $T^\mu_\rho (x)$, to transform it  to a new frame, whereas the RHS will need factors of $T^\mu_\rho(x_0)$. Thus the equation holds in any coordinate system if the RHS is multiplied by factors of $T^\mu _\rho(x) (T^{-1})^\rho _\sigma (x_0)$. 

We can apply this to the metric tensor $g_{\mu\nu}^R$. Thus the equation becomes:
\[
\bar g_{\rho \sigma}(Y) = \bar g_{\rho \sigma}(x_0) - {1\over 3}
 \bar R_{\sigma\mu \rho \nu}(x_0) 
 y^\mu y^\nu +...
\]
In particular if $R=0$ we get $\bar g_{\rho \sigma}(Y)= \bar g_{\rho \sigma}(x_0)$, which is just the statement that
in RNC, a flat metric is constant. We can transform to arbitrary coordinates and get the expected result, that the (flat) metric is no longer constant: 
\[
g_{\mu \nu}(x)= (T(x)T^{-1}(x_0))^\rho_\mu(T(x)T^{-1}(x_0))^\sigma_\nu  g_{\rho\sigma}(x_0)
\]

Now consider the kinetic term $g_{\mu \nu}^R(X) \p_zX^\mu \p_z X^\nu$. This is a scalar, so we will first write it in the RNC
coordinates as
 \[
 g_{\mu \nu}^R(Y) \p_zY^\mu \p_\zb Y^\nu = \bar g_{\mu \nu}^R(x_0+y) \p_z y^\mu \p_\zb y^\nu
 \]
 \[
 =( \bar g_{\mu \nu}^R(x_0)  - {1\over 3}
 \bar R^R_{\sigma\al \rho \beta}(x_0) 
 y^\al y^\beta +...) \p_z y^\mu \p_\zb y^\nu
 \]
 Let us once again specialize to the case where the metric $g^R$ is flat. Then we have
\be   \label{KEbar}
 = \bar g_{\mu \nu}^R(x_0)  \p_z y^\mu \p_\zb y^\nu
 \ee

 Since this expression is a scalar (remember that $y^\mu$ is a geometric object and has a meaning independent of coordinates), the kinetic term  is the same in any coordinate system, so we can remove the bars: 
 \be   \label{KE}
 =
 g_{\mu \nu}^R(x_0)  \p_z y^\mu \p_\zb y^\nu
 \ee
 $y^\mu$ will change to a transformed vector $y'^{\mu}$, but it is a variable of integration, so we have retained the same notation.
Now let us turn to the interaction term
\[
\tilde h_{\mu\nu}(X)\p_z X^\mu \p_\zb X^\nu
\] 
This is a scalar and we can write it directly in RNC to get
\[
=\bar {\tilde h}_{\mu\nu}(Y)\p_z Y^\mu \p_\zb Y^\nu = \bar {\tilde h}_{\mu\nu}(x_0+y)\p_z y^\mu \p_\zb y^\nu
\]
We can Taylor expand using (\ref{Taylor}) and setting background curvature to zero,  to get
\[
=(\bar{\tilde h}_{\mu\nu}(x_0) + y^\rho \nabla^R_\rho \bar{\tilde h}_{\mu\nu}(x_0) + {1\over 2!}y^\rho y^\sigma \nabla ^R_\sigma \nabla^R_\rho \bar{\tilde h}_{\mu\nu}(x_0) +...)\p_z y^\mu \p_\zb y^\nu
\]
Once again each term in the expansion is a scalar and has the same value in any coordinate system, so we can remove the bars:
\[
=({\tilde h}_{\mu\nu}(x_0) + y^\rho \nabla^R_\rho {\tilde h}_{\mu\nu}(x_0) + {1\over 2!}y^\rho y^\sigma \nabla ^R_\sigma \nabla^R_\rho {\tilde h}_{\mu\nu}(x_0) +...)\p_z y^\mu \p_\zb y^\nu
\]

We can also write the above as
\[
\int dk~\tilde h_{\mu\nu}(x_0,k_0)e^{ik_0.y}\p_z y^\mu \p_\zb y^\nu
\]
where  each power of $k_\mu$ is a background covariant derivative $\nabla_\mu^R$. Since curvature tensors are all zero, the covariant derivatives commute, so this is consistent.
Similar expansions have to be made for the massive modes. This involves some subtlety because loop variables are necessary. This is discussed in the section 5.
Now we have  manifestly background covariant kinetic and interaction terms and we can apply the ERG.

\subsection{Free Equation for Graviton}

Let us now write down the free equation for $h_{\mu \nu}$. We set $\lan K_{1;\bar 1\mu}\ran=S_{\mu}=0$ and $\lan \kim k_{\bar 1\nu}\ran=\tilde h_{\mu\nu}= (h_{\mu \nu}-h_{\mu\nu}^R)$. We also replace
$\kom \rightarrow \nabla^R_\mu$ in (\ref{grav}). This gives
\[
[-\ko^2 \kim k_{\bar 1\nu} + \ko .\ki \kom k_{\bar 1\nu} + \ko. k_{\bar 1}\kon \kim - k_1.k_{\bar1} \kom \kon]Y_{1;0}^\mu Y_{0;\bar 1}^\nu=0
\]
\be
\Rightarrow -(\nabla^R)^ 2 (h_{\mu\nu}-h_{\mu\nu}^R)(x_0) + \nabla^R_\mu \nabla^{R\rho} (h_{\rho \nu} -h_{\rho \nu}^R)(x_0)+ \nabla^R_\nu\nabla^{R\rho} (h_{\rho \nu}- h_{\rho \nu}^R)(x_0)=0 
\ee
The argument of the fields, $x_0$, is explicitly indicated - this is a tensor equation at $x_0$.
 We have set $K_{1;\bar 1}=0$ so the  K-constraint is $\ko.K_{1;\bar 1}=\ki.\kib=0$ - see the next paragraph.

\subsection{The dilaton and K-constraint for the graviton}

At this point we can observe the following: Parametrizing $h_{\mu\nu}^R$ by $\xi_{(\mu ,\nu)}$ (for small $h^R$) we can write, integrating by parts on either $z$ or $\zb$:
\[
\xi _\mu \p_z \p_\zb X^\mu = -\p_\zb(\xi_\mu)\p_zX^\mu = - \xi_{\mu,\nu}\p_\zb X^\nu \p_z X^\mu =- \xi_{\mu,\nu}\p_\zb X^\mu \p_z X^\nu
\] 
\[
=-\hf \xi_{(\mu,\nu)}\p_\zb X^\nu \p_z X^\mu 
\]
Thus adding the reference metric term is equivalent, as far as symmetry properties are concerned, to adding the mixed derivative 
loop variable term $K_{1;\bar 1}^\mu \p_z\p_\zb X^\mu e^{i\ko.X}$! Thus in the present construction we do not need it and we can set
$K_{1;\bar 1}^\mu=0$.

The K-constraint for the graviton now reads as
\be	
K_{1;\bar1}.\ko = k_1.k_{\bar 1} = k_1.k_{\bar 1}+ q_1 q_{\bar 1}=0
\ee
where we have separated out the $D+1$th coordinate. $\lan q_1 q_{\bar 1}\ran = \Phi_D$ is the dilaton.  So the constraint reads
\be	\label{dilaton}
h^{R\mu}_{~~\mu}- h^\mu_{~\mu} + \Phi_D=0
\ee
Index contractions are done using $g_{\mu\nu}^R$. This equation is gauge covariant under $\delta_T+\delta_G$. It relates the trace of the metric to the dilaton. We remind the reader that in the old covariant formulation of string theory, the role of the dilaton is played by the trace of the metric. The physical graviton is transverse and traceless.
In the present formalism, we have both the trace of the metric and a dilaton. They are related by the constraint but the constraint does
not fix gauge because it is gauge covariant.  

\subsection{Interactions and Gauge Invariant Field Strength}

The expression for the gauge invariant field strength at level 2 is
 \be	\label{fldstrlevel2}
 -\kor k_{1\mu} k_{\bar 1\nu} + k_{1\rho} \kom k_{\bar 1\nu} + k_{\bar 1\rho} k_{1\mu} \kon - K_{1;\bar 1\rho} \kom \kon 
 \ee 
In terms of space time fields this is 

\[
\Big( -\nabla^R _\rho (h_{\mu \nu} -h_{\mu \nu}^R+ B_{\mu \nu})+ \nabla^R _\mu (h_{\rho \nu}-h_{\rho \nu}^R+B_{\rho \nu})+ \nabla^R _\nu (h_{\mu \rho}-h_{\mu \rho}^R+ B_{\mu \rho})\Big)
\]
\[
=\Big( -\nabla^R _\rho (h_{\mu \nu} -h_{\mu \nu}^R))+ \nabla^R _\mu (h_{\rho \nu}-h_{\rho \nu}^R)+ \nabla^R _\nu (h_{\mu \rho}-h_{\mu \rho}^R\Big)+\Big(-\p^\rho B_{\mu \nu}-\p^\mu B_{\nu \rho }-\p ^\nu B_{ \rho\mu})\Big)
\]
\be    \label{modspin2}
G_{\rho\mu\nu}\equiv\Big( -\nabla^R _\rho (h_{\mu \nu} -h_{\mu \nu}^R))+ \nabla^R _\mu (h_{\rho \nu}-h_{\rho \nu}^R)+ \nabla^R _\nu (h_{\mu \rho}-h_{\mu \rho}^R\Big)-H_{\rho \mu \nu}
\ee
where $H=dB$ is a gauge invariant 3-form field strength for $B$. Thus $G$ is a tensor and one can easily write down interaction terms from the ERG in terms of $G$ and other modes. One such equation is given below (\ref{CovSpin4}).

We thus get background gauge covariant interacting equations for $h_{\mu \nu}$. One can legitimately ask whether we can identify
$h$ as the graviton since we do not see Einstein's equation here. Einstein's equation for $h$ (without $h^R$) is expected to emerge after integrating out all the massive modes. {\em The fact that we have a massless symmetric rank 2 tensor with the correct gauge transformation properties guarantees that we will obtain Einstein's equations as the low energy limit of the ERG.} 

As explained in section 4.2,
the background covariance also helps to ensure that the final equation has to be generally covariant: This is because the background metric, $h^R$,  is completely arbitrary and, at least formally, cannot appear in the final answer as expressed by (\ref{hR}).  Then the manifest background covariance guarantees the full covariance of the final result. The background covariance of the massive modes is discussed in the next section.
In  Appendix C we give a toy model illustrating some of these points using Yang-Mills theory. 

\section{Massive Field Equations and some Speculations}

\subsection{Massive Field Equations} 
 The above technique for achieving covariance should go through for massive modes as well, provided they are tensors. This requires a modification of what we mean by "GCT" and what our manifold is. We outline the arguments although the details have not been worked out yet.
 
The action of $\delta_T+ \delta _G$ on massive modes is also quite simple, if $\delta _T$ is defined suitably. We use loop variables as usual, so instead of $\p_z^n
X^\mu$ 
we have $Y_n^\mu$. This ensures the invariance under $\delta _G$ as explained in Section 3. To understand the action of $\delta _T$
we notice that the final equation of motion has products of $Y_n^\mu$ and $Y_\mb^\nu$. We need to define what we mean by $\delta _T$ on  $Y$. We will take it to be $Y^\mu\rightarrow Y^{'\mu}(Y)$. This is not a consequence of the GCT on $X$: $X^\mu\rightarrow X^{'\mu}(X)$, since the
$Y$ is a very particular combination of $X$ and its derivatives. 

We will assume then that our differential manifold is labelled by $Y^\mu$ and assume that diffeomorphisms of this manifold are a symmetry. Thus $\delta _{GCT}$ acts on $Y$ rather than $X$ \footnote{Note that if we set  $\xn=0$, then $Y=X$.}. We can easily see that it is a symmetry of the EOM.
It is easy to see that $Y_n^\mu$ transforms as a tensor (more specifically, a  vector) under :
\be
Y^\mu (z)\rightarrow Y^{'\mu}(z)~~~~\Rightarrow {\p Y^\mu\over \p \xn} \rightarrow {\p Y^{'\mu}\over \p Y^\rho}{\p Y^\rho \over \p \xn} 
\ee 
(This is to be contrasted with the more complicated transformation of $Y^\mu_{[n];[\mb]}$. However the final equations for physical modes do not involve these vertex operators, so we do not have to worry about them at this stage. We will touch upon this in the next section.) Thus the massive modes transform as ordinary tensors under $\delta _T$. So the equations of motion are manifestly invariant (or covariant, if we remove the vertex operators multiplying the equations), provided we assign the usual tensorial properties to the fields. Contraction of indices is done using $g_{\mu\nu}^R$. Ordinary derivatives can be replaced by covariant derivatives: $\p _\mu \rightarrow \nabla^R_\mu=\p_\mu + \Gamma ^R_\mu$ in the usual way. Since the curvature is zero, the covariant derivatives commute:
$[\nabla^R_\mu ,\nabla^R_\nu]=0$, and the gauge invariance under $\delta _G$ is preserved. Thus in   (\ref{3243}), we simply replace $\ko$ by $\nabla^R$ and raise and lower indices with $g_{\mu\nu}^R$.
\[
\int dz~\dot G(z,z;\tau) (-(\nabla^R) ^2 S_{\mu \nu \rho\sigma} + \nabla^{R\la} \nabla^R_{(\mu}S_{\nu) \la \rho \sigma} + \nabla^{R\la} \nabla^R_{(\sigma} S_{|\mu \nu \la|\rho)}
\]
\be	\label{CovSpin4}
-\nabla^R_{\mu}\nabla^{R}_{\nu} S_{\la ~\rho \sigma}^{~\la} -\nabla_{R\rho} \nabla_{R\sigma} S_{\mu \nu \la}^{~~~~\la}+\nabla^R_{(\sigma|} \nabla^R_{(\nu}S_{ \mu)\la~|\rho)}^{~~~~\la})
+ \int dz ~\int dz'~\dot G(z,z';\tau) G^\la _{~~\mu \rho} G_{\la \nu \sigma} +...=0
\ee

 This procedure gives covariant and gauge invariant equations for all the modes. However a detailed understanding of the meaning of   coordinate transformations on $Y^\mu$ (as against $X^\mu$) is still lacking.

\subsection{Speculations on space time interpretation of massive gauge transformations}

The transformation $\delta X^\mu (z) = -\eps ^\mu(X(z))$  on massive vertex operators gives a transformation that is more complicated than that of $\p_z X^\mu$. For instance,
\[
\p_z^2X^\mu(z)\rightarrow \p_z^2X^{'\mu}(z) = \p_z(\p_z X^\mu - \eps^\mu_{~,\rho}\p_zX^\rho)
\]
\be
=\p_z^2X^\mu- \eps^\mu_{~,\rho\sigma}\p_zX^\rho \p_zX^\sigma- \eps^\mu_{~,\rho}\p_z^2 X^\rho
\ee
 The same transformation rule is obtained if we consider $\p^2 Y^\mu \over \p x_1^2$ with a transformation $\delta Y^\mu (z) = -\eps ^\mu(Y(z))$:
 \[
{ \pp Y^\mu \over \p {x_1}^2}\rightarrow {\pp Y^{'\mu} \over \p {x_1}^2} =
  \p_{x_1}(\p_{x_1}Y^\mu - \eps ^\mu _{~,\rho}\p_{x_1}Y^\rho)
 \]
 \be
= \p_{x_1}^2Y^\mu - \eps ^\mu _{~,\rho\sigma }\p_{x_1} Y^\rho \p_{x_1} Y^\sigma -\eps ^\mu _{~,\rho} \p_{x_1}^2 Y^\rho
\ee
The transformation has a term that mixes different tensor structures (but at the same mass level), in addition to the usual tensorial
transformation.
This is to be contrasted with the transformation of $Y_2^\mu$:
thus $Y_{1,1}^\mu$ and $Y_2^\mu$ although they are equal, transform differently.  In the loop variable formalism as described in Section 3 for closed strings and
I and II for open strings, the final equations involve only $Y_n^\mu$, which also transforms as a vector. But the intermediate stages involve $Y_{[n]}^\mu$ ($[n]$ stands for a partition of $n$), which has a more complicated transformation. Indeed for closed strings we also need $Y_{1;\bar 1}^\mu$, which also does not transform as a simple vector. These transformations mix all the tensors at a given mass level. The question arises as to whether
these more complicated transformations have any connection with the massive gauge transformations, just as the GCT is related to massless graviton transformations. We give an argument that is suggestive of such a connection but is not definitive. It involves
the transformation of terms involved in regularizing the world sheet theory.

\subsection{Regularization and Massive Modes}

In order to derive a Renormalization Group we need to regularize the theory. Thus our starting point is
\[
\int dz_1 dz_2~ \eta_{\mu \nu}f(z_1-z_2,a)\p_z X^\mu(z_1)\p_\zb X^\nu(z_2) \equiv \int dz~\eta_{\mu\nu}\sum_{n,m}f_{n,m}(a)\p_z^nX^\mu(z)^\mu \p_\zb^m X^\nu(z)
\]
In (world sheet) momentum space this would correspond to 
\[
\int dp~\sum_{n,m}f_{n,m}X^\mu(p)p^n\bar p^m X^\nu(-p)\eta_{\mu \nu}
\]
which would cutoff the high momentum region in the loop integrals. 

Let us now consider the effect of one of these terms. We will write in terms of $Y$:
\be
c_2 \eta_{\mu \nu} \p_{x_1}^2Y^\mu\p_{\bar x_1}^2Y^\nu
\ee
Let us transform $Y_{1,1}^\mu$.
\be
c_2 \eta_{\mu\nu}[( \p_{x_1}^2Y^\mu - \eps ^\mu _{~,\rho\sigma }\p_{x_1} Y^\rho \p_{x_1} Y^\sigma -\eps ^\mu _{~,\rho} \p_{x_1}^2 Y^\rho)\p_{\bar x_1}^2Y^\nu -(\eps^\nu_{~,\rho} \p_{\bar x_1}^2 Y^\rho +  \eps ^\nu_{~,\rho \sigma}\p_{\bar x_1}Y^\rho\p_{\bar x_1}Y^\sigma) \p_{x_1}^2 Y^\mu]
\ee
"c.c" stands for the transformation of $\p_{\bar x_1}^2Y^\nu$. We will not write this out explicitly, in order to keep the algebra simple.
Thus in analogy with what was done for the kinetic term , let us introduce the reference fields $h^{2R}_{\mu\nu}$ and $h^{2R}_{\mu \rho \nu}$ by adding and subtracting the terms
\[ c_2  h^{2R}_{\mu\nu} \p_{x_1}^2 Y^\mu\p_{\bar x_1}^2Y^\nu+
c_2 h^{2R}_{ \rho \sigma \mu}(\p_{x_1} Y^\rho \p_{x_1} Y^\sigma \p_{\bar x_1}^2Y^\mu +\p_{\bar x_1} Y^\rho \p_{\bar x_1} Y^\sigma \p_{ x_1}^2Y^\mu) 
\]

Thus if we assume that $h^{2R}_{\mu \nu}$ and $h^{2R}_{\mu \nu \rho}$ transform as tensors and additionally have the variation
(defining as before $\eps_\mu \equiv \eta_{\mu \nu} \eps^\nu $)
\[
\delta h^{2R}_{\mu \nu} = \eps _{(\nu,\mu)};~~~\delta h^{2R}_{\rho\sigma\mu}= \eps _{\mu,\rho \sigma}
\]
then this term is invariant. 

the interaction Lagrangian now contains the same terms with the opposite sign:
\be	\label{deltaL}
\Delta L=-(c_2  h^{2R}_{\mu\nu} \p_{x_1}^2 Y^\mu\p_{\bar x_1}^2Y^\nu+
c_2 h^{2R}_{ \rho \sigma \mu}(\p_{x_1} Y^\rho \p_{x_1} Y^\sigma \p_{\bar x_1}^2Y^\mu +\p_{\bar x_1} Y^\rho \p_{\bar x_1} Y^\sigma \p_{ x_1}^2Y^\mu)) \ee

We already have a term
\[
 K_{1,1;\bar 1,\bar 1\mu} \p_{x_1}^2\p_{\bar x_1}^2 Y^\mu \e
\]
Integrating by parts on $x_1$ we get
\be	\label{K1111}
K_{\mu1,1:\bar 1,\bar 1}(-\hf \ko ^\sigma \kor( \p_{x_1}Y^\sigma \p_{x_1}Y^\rho \p_{\bar x_{1}}^2Y^\mu +\p_{\bar x_1}Y^\sigma \p_{\bar x_1}Y^\rho \p_{ x_{1}}^2Y^\mu)+ i\kor \p_{x_1}^2Y^\rho \p_{\bar x_1}^2Y^\mu)\e
\ee
 Comparing  (\ref{deltaL}) with (\ref{K1111}) we see that we can identify $\kom\kon K_{\rho1,1;\bar 1,\bar 1}$ with $2c_2h^{2R}_{\mu\nu\rho}$ and $\kom K_{\nu1,1;\bar 1,\bar 1}$ with $2c_2 h^{2R}_{\mu\nu}$ and also the gauge transformation $K_{\nu1,1;\bar 1,\bar 1}\rightarrow
 K_{\nu1,1;\bar 1,\bar 1}+ \la_1 K_{\nu1;\bar 1,\bar 1}+ \bar \la_1K_{\nu1,1;\bar 1}$ with that of $h^{2R}$ if we identify $\lan \la_1 K_{\nu1;\bar 1,\bar 1}+ \bar \la_1 K_{\nu 1,1;\bar 1}\ran =2\eps _\nu$.
 
 This suggests that just as for the graviton, the extra terms can be associated with variations of the regulator kinetic terms. We find this extremely interesting because the same argument that allowed us to have a massless graviton can be applied here, but now to the genuinely massive modes: they can be massless as well. Thus somehow at the scale of the cutoff, these modes could be massless.
 Alternatively, the physical interpretation could be that this is another consistent phase of string theory.  
 
 This analysis is clearly incomplete because the parameter $\eps$ is the same as for the GCT considered in Section 4. Whereas the gauge parameter $\lan \la_1 K_{\nu1;\bar 1,\bar 1}+\bar \la_1 K_{\nu 1,1;\bar 1}\ran $ should be independent. This is because we have restricted $\eps$ to be only a function of $Y$. More generally it could be a function of all the $Y_n$. This will generate transformations that mix different mass levels and tensor structures. We leave this as an open question.
 
 \subsection{Complexification of Space time coordinates?}
We identified
\be	\label{eps}
\lan \la_1 k_{\bar 1\mu} + \bar \la _1 k_{1\mu}\ran = \eps_\mu = -\eta_{\mu\nu}\delta X^\nu
\ee

Also
\be	\label{Lamb}
\lan \la_1 k_{\bar 1\mu} - \bar \la _1 k_{1\mu}\ran =\Lambda_{1\mu}
\ee
is the gauge transformation parameter of the antisymmetric tensor.
Comparing the two strongly suggests that $X^\mu$  be thought of as the real part of a complex coordinate and that
(\ref{Lamb})  be identified with the variation of the imaginary part. That space time coordinates could at some level be complex was suggested in \cite{WTop,AW}. 

 This certainly requires further investigation.
 \section{Summary and Conclusions}
 
 In this paper we have extended the construction in I and II, of a gauge invariant ERG for open strings to closed strings. The salient features are the following: 
 \begin{enumerate}
 \item
 The construction is restricted to flat geometry for simplicity. Thus the graviton is a perturbation about flat space. Nevertheless
 we introduce a reference metric, so that arbitrary coordinate transformations can be made and we need not restrict the flat metric to be of the form $\eta_{\mu \nu}$. (We chose the reference metric to be of zero curvature in order to simplify the results. This can be relaxed.) The final result is an EOM for closed strings that has general coordinate invariance in the sense of background field theory: Transform coordinates and the background reference metric (of zero curvature) and transform all other fields as tensors. The physical graviton
$h_{\mu\nu}$ occurs in combination with $h_{\mu\nu}^R$ in the form $\tilde h = h-h^R$ which is a tensor. The original Abelian gauge
invariance is embedded in the general coordinate transformation of $h$. 
  \item In order to apply the technique of I and II to closed strings it was found necessary to include in the intermediate stages
 of the calculation, vertex operators involving $\p_z \p_\zb X$. This is also expected for independent reasons: Euclidean world sheet regularization breaks the holomorphic factorization. Thus on the scale of the cutoff one should expect to add such terms. They decouple in the continuum limit. These fields are not there in the BRST formalism.
 \item
 The EOM for physical vertex operators are gauge invariant under the natural generalization of the open string gauge transformations to closed strings. These are Abelian. While this is expected for open strings, this is not expected for closed strings. The resolution of this
 lies in another problem with this construction: as it stands the graviton cannot be massless. Gauge invariance requires an auxiliary field
 which can be written down in terms of loop variables only if the graviton is massive. Both these problems are resolved by extending the symmetry transformations to act on the space time coordinates as well. This introduces a Christoffel connection  term for a background metric, that obviates the need for an independent  auxiliary field. This makes the graviton massless and also makes the symmetry that of general coordinate transformations, as is appropriate for a theory of gravity.
 \item
 The gauge transformations of massive fields continue to be Abelian, if one considers only those equations connected with
 the physical vertex operators. However if one considers the vertex operators at intermediate stages of the calculation, there are
 more complicated transformations. There is some preliminary indication of a more elaborate space time interpretation that mixes different tensor structures and mass levels.
 \item
 The EOM are quadratic as expected from an ERG. This is different from the BRST formalism. The can possibly be attributed to the background field formalism. We have seen that the extra fields involving mixed derivatives is related to the background metric. 
  \end{enumerate}
 
 There are many conceptual and technical questions that need to be answered. We list a few:
 \begin{enumerate}
 \item
 The loop variable formalism is formally written with 
 an extra dimension. The massive equations are obtained by dimensional reduction. This is in principle straightforward, but the details need to be worked out. 
 \item
 In the case of open strings it was found at the first and second massive levels,  that the gauge transformations and constraints can be mapped on to those of the "Old Covariant" formalism (only) in D=26 and with the correct mass levels. Thus we expect that in the critical dimension the S-matrix of this theory should coincide with that of string theory. Some arguments for the equality of the S-matrix were given in \cite{BSOC}. This needs to be made more rigourous and a similar analysis needs to be done for the closed string.
 \item
 In dimensions other than 26, this seems to be a consistent classical theory of massive higher spins interacting with gravity. This is because the gauge symmetries are the same in all dimensions. Whether there are inconsistencies at the quantum level is an open question. These issues need to be sorted out.
 \item
 One may also expect a more direct connection to the critical dimension by studying the EOM of the dilaton \cite{CDMP}.  This involves  technical issues related to overall normalization of the partition function that we have not worried about in this paper.
 \item
 A flat background metric was chosen  to avoid the additional complication of  modifying the map from loop variables to space time fields for the massive modes.  Non flat background metrics can be chosen. In this case one has the option of setting $h_{\mu\nu}-h^R_{\mu\nu}=\tilde h_{\mu\nu} =0$ at any point. This will then  give equations of motion that are quadratic in the massive fields, but non polynomial in the graviton. This is also worth exploring.
 \item
 The connection between massive gauge transformations and space time coordinate transformations (or generalizations thereof) need to be worked out. The interplay between coordinate transformations $Y\rightarrow Y'$ and gauge transformations $Y\rightarrow Y+ {\p Y\over \p \xn}$  needs to be fully understood. 
 \item
 We have now  background gauge covariant equations of motion. The problem of constructing a gauge invariant action is unsolved (for open strings also).
 \end{enumerate}

{\bf Acknowledgements:} I would like to thank Ghanashyam Date, S. Kalyana Rama and Partha Mukhopadhyay for useful discussions.
\appendix

\renewcommand{\thesection}{\Alph{section}}
\renewcommand{\theequation}{\thesection.\arabic{equation}}

\section{Appendix: Free Equation}
\label{appena}
\setcounter{equation}{0}

The details of the calculation of the Level 2 (graviton) and Level 4 free equation are given here.

We have to evaluate the second derivative, which is given by the action of a functional derivative on (\ref{FnlDer}):
\[
=\int dz'\int dz''~\dot G(z',z'')~
\]
\[\eta^{\mu\nu}\int du~[{\p\over \p Y^\nu(u)}\delta(u-z') +[{\p\over \p x_1}\delta(u-z')]{\p \over \p Y^\nu_{1;0}(u)}+
\]\[
[{\p\over \p \bar x_1}\delta(u-z')]{\p \over \p Y^\nu_{0;\bar1}(u)}+[{\p\over \p x_2}\delta(u-z')]{\p \over \p Y^\nu_{2;0}(u)}+\]
\[[{\p\over \p \bar x_2}\delta(u-z')]{\p \over \p Y^\nu_{0;\bar 2}(u)}+
 [{\pp\over \p x_1\p \bar x_1}\delta(u-z')]{\p \over \p Y^\nu_{1;\bar 1}(u)}+...]\]
 \[
 \Big\{ \underbrace{{\p {\cal L} [Y(u),Y_{n,\mb}(u)]\over \p Y^\mu(u)} \delta (u-z'')}_I-\underbrace{\p_{x_1}{\p {\cal L} [Y(u),Y_{n;\mb}(u)]\over \p Y_{1;0}^\mu(u)} \delta (u-z'')}_{II}
\]
\be - \underbrace{\p_{\bar x _1} {\p {\cal L}[Y(u),Y_{n;\mb}(u)] \over \p Y_{0;\bar 1} ^\mu(u) }
\delta (u-z'')}_{III} +  \underbrace{\p _{ x_{1}}\p_{\bar x_{1}}{\p {\cal L}[Y(u),Y_{n,\mb}(u)] \over \p Y_{1,\bar 1}^\mu (u) }
 \delta (u-z'')}_{IV} \Big\}
\ee

Let us evaluate the action of the derivatives on each of the four terms labeled I,II,III and IV. We can reduce the number of independent terms to be evaluated by realizing that the result has to be symmetric in $z'\leftrightarrow z''$ and also that for every term, there is also a corresponding complex conjugate term. (Our notation is: $\xn$ refers to $u$, $\xn '$ refers to $z'$ and $\xn ''$ refers to $z''$. Thus for instance, ${\p \delta(u-z')\over \p \xn}=-{\p \delta(u-z')\over \p \xn '}$)
\begin{enumerate}
\item 
\[
\int du~\eta^{\mu \nu}{\p\over \p Y^\nu(u)}{\p {\cal L} [Y(u),Y_{n,\mb}(u)]\over \p Y^\mu(u)} \delta(u-z')\delta (u-z'')
\]
\be=
-\ko^2 {\cal L} (z')\delta(z'-z'')
\ee
\item
\[
\int du~\eta^{\mu \nu}\Big([{\p\over \p x_1}\delta(u-z')]{\p \over \p Y^\nu_{1;0}(u)}{\p {\cal L} [Y(u),Y_{n,\mb}(u)]\over \p Y^\mu(u)} \delta (u-z'')\Big) +\Big( z'\leftrightarrow z''\Big) 
\]
\[=~\eta^{\mu \nu}\Big(-{\p\over \p x'_1}[\delta(z''-z')i\ko.iK_{1;0}{\cal L}[z'']] \Big) +\Big( z'\leftrightarrow z''\Big)
\]
\[=\eta^{\mu \nu}\Big(-[{\p\over \p x'_1}+{\p\over \p x''_1}][\delta(z''-z')i\ko.iK_{1;0}{\cal L}[z'']] \Big)
\]

We restore the integrals over $z',z''$, and use $G(z',z'')= \lan Y(z')Y(z'')\ran$ and integrate by parts on $x',x''$ to get
\[{d\over d ln~ a}\int dz'dz''~({\p\over \p x'_1}+{\p\over \p x''_1})\lan Y(z')Y(z'')\ran [\delta(z''-z')i\ko.iK_{1;0}{\cal L}[z'']]=
\]
\[=
{d\over d ln~ a}\int dz'~[{\p\over \p x'_1}\lan Y(z')Y(z')\ran] i\ko.iK_{1;0}{\cal L}[z']]\]
\[
=-\int dz'~\dot G(z',z') i\ko.iK_{1;0}{\p\over \p x'_1}[{\cal L}[z']]
\]
In the last step we have integrated by parts again.

Finally we can add the complex conjugate to obtain:
\be
=\int dz'~\dot G(z',z')\Big( \ko.K_{1;0}{\p\over \p x'_1}[{\cal L}[z']]+\ko.K_{0;\bar 1}{\p\over \p \bar x'_1}[{\cal L}[z']]\Big)
\ee
\item
\[
\eta ^{\mu \nu}\int dz'\int dz''~\dot G(z',z'')\int du~[{\p\over \p x_1}\delta(u-z')][{\p\over \p x_1}\delta(u-z'')]{\pp{\cal L}[u]\over \p Y^\mu_{1;0}(u)\p Y^\nu_{1;0}(u)} 
\]
\[=
{d\over d ln~ a}\eta ^{\mu \nu}\int dz'\int dz'' \lan Y_{1;0}(z')Y_{1;0}(z'')\ran \delta(z'-z''){\pp{\cal L}[z']\over \p Y^\mu_{1;0}(z')\p Y^\nu_{1;0}(z')} 
\]
\[=
{d\over d ln~ a}\eta ^{\mu \nu}\int dz'[\hf ({\pp \over \p x_1^{'2}}-{\p \over \p x'_{2}})\lan Y(z')Y(z')\ran]{\pp{\cal L}[z']\over \p Y^\mu_{1;0}(z')\p Y^\nu_{1;0}(z')} 
\]
\[=
\eta ^{\mu \nu}\int dz'\dot G(z',z')\hf ({\pp \over \p x_1^{'2}}+{\p \over \p x'_{2}}){\pp{\cal L}[z']\over \p Y^\mu_{1;0}(z')\p Y^\nu_{1;0}(z')} 
\]
\be
=-\int dz'\dot G(z',z') K_{1;0}.K_{1;0}\hf ({\pp \over \p x_1^{'2}}+{\p \over \p x'_{2}}){\cal L}[z']
\ee
\item The complex conjugate is:
\be
-\int dz'\dot G(z',z') K_{0;\bar 1}.K_{0;\bar1}\hf ({\pp \over \p \bar x_1^{'2}}+{\p \over \p \bar x'_{2}}){\cal L}[z']
\ee
\item
\[
\int dz' dz'' ~\dot G(z',z'')\Big( \eta^{\mu \nu}\int du~[{\p\over \p x_2}\delta(u-z')]\delta(u-z''){\pp{\cal L}[u]\over \p Y^\mu_{2;0}(u)\p Y^\nu(u)} ~+~z'\leftrightarrow z'' \Big)
\]
\[
= \int dz' dz'' ~\dot G(z',z'')\Big( \eta^{\mu \nu}~[-{\p\over \p x'_2}\delta(z''-z')]{\pp{\cal L}[z'']\over \p Y^\mu_{2;0}(z'')\p Y^\nu(z'')} ~+~z'\leftrightarrow z'' \Big)
\]
\[
= \int dz' dz'' ~{d \over d ln~a} [{\p\over \p x'_2}+ {\p\over \p x''_2}]G(z',z'')\Big( \eta^{\mu \nu}~[\delta(z''-z')]{\pp{\cal L}[z'']\over \p Y^\mu_{2;0}(z'')\p Y^\nu(z'')} \Big)
\]
\be
= \int dz'  ~\dot G(z',z')\Big(K_{2;0}.\ko~({\p\over \p x_2}{\cal L}[z'']) \Big)
\ee
\item
Complex conjugate gives:
\be
= \int dz'  ~\dot G(z',z')\Big(K_{0;\bar 2}.\ko~({\p\over \p \bar x_2}{\cal L}[z'']) \Big)
\ee
\item
\[
\int dz' dz'' ~\dot G(z',z'')\Big( \eta^{\mu \nu}\int du~[{\pp\over \p x_1 \p \bar x_1}\delta(u-z')]\delta(u-z''){\pp{\cal L}[u]\over \p Y^\mu_{1;\bar 1}(u)\p Y^\nu(u)} ~+~z'\leftrightarrow z'' \Big)
\]
\[ 
={d\over d~ln~a}\int dz' dz'' ~ \lan Y (z')_{1;\bar 1} Y(z'')+Y (z') Y_{1;\bar 1}(z'')\ran \Big( \eta^{\mu \nu} \delta(z'-z''){\pp{\cal L}[z']\over \p Y^\mu_{1;\bar 1}(z')\p Y^\nu(z')} \Big)
\]
\be  \label{8}
={d\over d~ln~a}\int dz' dz'' ~ \lan Y (z')_{1;\bar 1} Y(z'')+Y (z') Y_{1;\bar 1}(z'')\ran \Big(( iK_{1;\bar 1}.i\ko ) \delta(z'-z''){\cal L}[z'] \Big)
\ee
\item
\[
\int dz' dz'' ~\dot G(z',z'')\Big( \eta^{\mu \nu}\int du~[{\p\over \p \bar x_1 }\delta(u-z')][{\p \over \p x_1}\delta(u-z'')]{\pp{\cal L}[u]\over \p Y^\mu_{1;0}(u)\p Y^\nu_{0;\bar 1}(u)} ~+~z'\leftrightarrow z'' \Big)
\]
\be \label{9}
=
{d\over d~ln~a}\int dz' dz'' ~ \lan Y_{0:\bar 1}(z')Y_{1;0}(z'')+Y_{0:\bar 1}(z'')Y_{1;0}(z')\ran\Big( \delta(z'-z'')(iK_{1;0}.iK_{0;\bar 1}){\cal L}[z'] \Big)
\ee
\end{enumerate}
(\ref{8}) and (\ref{9}) can be added to give 
\[
=
{d\over d~ln~a}\int dz' ~ [{\pp\over \p x'_1\p \bar x'_1}\lan Y(z')Y(z')\ran]\Big( (iK_{1;0}.iK_{0;\bar 1}){\cal L}[z'] \Big)
\]
\be =
\int dz' ~ \dot G(z',z')\Big( (iK_{1;0}.iK_{0;\bar 1})[{\pp\over \p x'_1\p \bar x'_1}{\cal L}[z']] \Big)
\ee
{\em provided} the following constraint is imposed :
\be  \label{Kconstraint}
K_{1;0}.K_{0;\bar 1}~ {\cal L}= K_{1;\bar 1}.\ko ~ {\cal L}
\ee
The constraint is gauge covariant since both sides have identical gauge transformation properties.
Since $K_{1;\bar 1}$ is an auxiliary field (i.e. not physical) we are free to impose this constraint. In fact since $K_{1;\bar 1}.\ko$
contains $q_{1;\bar 1} q_0$  (for  $q_0\neq 0$) , this can be treated as an algebraic constraint on $q_{1;\bar 1}$.

The massless case (Graviton) is discussed in Section 3 and Section 4.

Similar constraints on $K_{n;\mb}$ occur at every level. We will refer to them as K-constraints. They are described in the next Appendix.

The terms calculated above are sufficient to extract the coefficient of the graviton multiplet vertex operators at level ($1;\bar 1$), $Y_{1;0}^\mu Y_{0;\bar 1}^\nu$ and the next massive level, $Y_{1;0}^\mu Y_{1;0}^\nu Y_{0;\bar 1}^\rho Y_{0;\bar 1}^\sigma \e$, a vertex operator in closed string theory at level $(2;\bar 2)$. We revert to the notation
$k_1=K_{1;0}, k_{\bar 1}=K_{0;\bar 1},...$ below.

We get for level $(1,\bar 1)$:
\be
[-\ko^2 \kim k_{\bar 1\nu} + \ko .\ki \kom k_{\bar 1\nu} + \ko. k_{\bar 1}\kon \kim - \ki.k_{\bar 1}\kom \kon]Y_{1;0}^\mu Y_{0;\bar 1}^\nu
\ee
At level $(2,\bar 2)$ for $Y_{1;0}^\mu Y_{1;0}^\nu Y_{0;\bar 1}^\rho Y_{0;\bar 1}^\sigma \e$ we get:

\[
-{1\over 4} \ko^2 (k_1 . Y_1)^2(k_{\bar 1}.Y_{\bar 1})^2 + \hf \ko.\ki (\ko . Y_1)(\ki .Y_1)(k_{\bar 1}.Y_{\bar 1})^2+\hf \ko.k_{\bar 1} (\ko . Y_{\bar 1})(k_{\bar 1} .Y_{\bar1})(k_{ 1}.Y_{1})^2
\]
\be   \label{EOMlevel4}
- {\ki .\ki\over 4} (\ko . Y_1)^2(k_{\bar 1}.Y_{\bar 1})^2 - {k_{\bar 1} .k_{\bar 1}\over 4} (\ko . Y_{\bar 1})^2(k_{ 1}.Y_{\ 1})^2 - \ki .k_{\bar 1} (\ko . Y_1)(\ko . Y_{\bar 1})(\ki .Y_1)(k_{\bar 1} .Y_{\bar1})
\ee

This can easily be seen to be gauge invariant under $\kim \rightarrow \kim + \li \kom$ and $k_{\bar 1\mu} \rightarrow k_{\bar 1\mu} +  \la _{\bar 1} \kom$, after using the tracelessness condition on the gauge parameter, $\li \ki .k_{\bar 1}=0=\li k_{\bar1}.k_{\bar 1}$ and the same for its complex conjugate.

\section{Appendix: K-constraints}
\setcounter{equation}{0}

We derive the K-constraints that occur in the free equations.

For the free part of the equation we do not need the individual $K_{[n]_i;[\mb]_j\mu}$. We can write ${\cal L}$ in terms of $Y_{n;\mb}^\mu$. Thus the coefficient of $Y_{n;\mb}^\mu$ is $\sum_{i,j}K_{[n]_i;[\mb]_j\mu} = \tilde  K_{n;\mb\mu}$ as defined in (\ref{intrel}).

The general case involves combining the following two terms:
\[
\int dz'\int dz''\dot G(z',z'') {\p \over \p \xn}{\p\over \p Y_m(u)}\delta(u-z') {\p \over \p x_\mb}{\p\over \p Y_\mb(u)}\delta(u-z''){\cal L}~+z'\leftrightarrow z''
\]
\be  \label{A12}
=
\int dz'{d\over d\tau}( [\lan Y_n (z')Y_\mb (z')\ran + \lan Y_\mb(z') Y_n(z')\ran](ik_n.ik_\mb) {\cal L}
\ee
and
\[
\int dz' \dot G(z',z'') {\pp\over \p \xn \p x_\mb} {\p\over \p Y_{n;\mb}}\delta(u-z'){\p\over \p Y}\delta(u-z''){\cal L} ~+z'\leftrightarrow z''
\]
\be	\label{A13}
=
\int dz' {d\over d\tau} [\lan Y_{n;\mb}(z') Y(z')\ran + \lan Y(z') Y_{n;\mb}(z')\ran](i\tilde K_{n;\mb}.i\ko) {\cal L} 
\ee

Now {\em if} 
\be	\label{Knm}
(ik_n.ik_\mb) {\cal L}=(i\tilde K_{n;\mb}.i\ko) {\cal L}
\ee
{\em then} we can combine the two terms, (\ref{A12}) and (\ref{A13}),   and write 
\[
\int dz' [{d\over d\tau} {\pp\over \p \xn \p x_{\mb}} \lan Y(z')Y(z')\ran] (-k_n.k_\mb){\cal L}
\]
\be
=\int dz' ~\dot G(z',z') (-k_n.k_\mb){\pp\over \p \xn \p x_{\mb}}{\cal L}
\ee

Since the $\tilde K_{n;\mb\mu}$ are made of the usual loop variables and no new degrees are involved, the K-constraints (\ref{Knm}) would seem
to reduce the number of independent degrees of freedom. However we also have the option of adding  {\em one new} loop variable $k_{n;\mb\mu}$ , (with $\mu$ chosen to be $D$, so we can call it $q_{n;\mb}$ )to $\tilde K_{n;\mb\mu}$ so that the constraint plays the role of determining this variable. $q_{n;\mb}$ should be defined to have the same gauge transformation as $\tilde K_{n;\mb\mu}$, viz: 

\[
q_{n;\mb}\rightarrow q_{n;\mb}+ \la _p q_{n-p;\mb} + \bar \la _p q_{n;\mb-p}
\]
Then the constraint does not affect the degrees of freedom count. 

We have 
\[
\tilde K_{n;\mb\mu} = \bar q_n k_{\mu\mb}+ \bar q_\mb k_{n\mu} - \bar q_n \bar q_\mb \kom
\]
\[
\tilde Q_{n;\mb}= {q_n\over q_0}q_\mb + {q_\mb\over q_0} q_n - {q_n\over q_0}{q_\mb\over q_0}q_0 +q_{n;\mb}={q_n q_\mb\over q_0}+q_{n;\mb}
\]

The constraint (\ref{Knm}) becomes
\[
q_{n;\mb}= k_n.k_{\mb} - \bar q_n k_\mb.\ko + \bar q_\mb k_n.\ko - \bar q_n \bar q_\mb \ko^2
\]
thus fixing $q_{n;\mb}$ in terms of the others. 

\section{Appendix: Utility of Background Fields: Toy Model}
\setcounter{equation}{0}

The exact ERG involves an infinite number of irrelevant coupling constants (massive modes), in addition to the marginal one, the graviton field. It is only after solving for all the
massive modes in terms of the graviton, and plugging back in, that one gets the full non polynomial equation for the graviton, which is the covariant Einstein equation 
along with covariant $\alpha ' $ corrections. Each equation by itself does not have the general covariance. However if we introduce
background fields, it is possible to make each equation by itself, generally covariant under a transformation which includes not only the usual transformations, but also a general coordinate transformation of the background field. The main advantage of this is that a symmetry, {\em which is very similar to the original symmetry}, is manifest throughout the calculation.  We illustrate this with a toy example.

Consider an $SU(2)$ Yang-Mills theory, with action.
\[
{1\over 4}Tr[F_{\mu \nu}F^{\mu\nu}]
\]
where $F_{\mu \nu}= \p_\mu A_\nu-\p_\nu A_\mu + i[A_\mu,A_\nu]$. Our conventions and notation are as follows:  $A_\mu = A_\mu^a \tau^a/2$ where $\tau^a$ are the Pauli matrices. If $\phi = \phi^a \tau^a/2$ is an adjoint field, it's gauge rotation is 
\[
\delta \phi =- i[\Lambda ,\phi]
\] Gauge rotation of $A_\mu$: 
\be   \label{gauge}
 \delta A_\mu =\p_\mu \Lambda +i[A_\mu ,\Lambda] 
\ee
Thus the covariant derivative is
$D_\mu \phi = \p_\mu \phi + i[A_\mu,\phi]$.

Now we write $A_\mu = W_\mu + \tilde Q_\mu$, where $W_\mu$ is some reference background field and $\tilde Q_\mu$ is the quantum field, which we will set to be $A_\mu - W_\mu$ at the end (by analogy with the $\tilde h=h-h^R$). So in fact nothing should depend on $W_\mu$.
 Then we can set 
\be \label{bgauge}
\delta W_\mu = \p_\mu \Lambda + i[W_\mu,\Lambda] ;~~~\delta \tilde Q_\mu = -i[\Lambda , \tilde Q_\mu]
\ee
The inhomogeneous term has been assigned to $W_\mu$. $\tilde Q$ transforms homogeneously. If we set $\tilde Q_\mu = A_\mu -W_\mu$, then this is still correct because the inhomogeneous terms cancels out.

\[
F_{\mu \nu} = (\p_\mu W_\nu-\p_\nu W_\mu + i[W_\mu,W_\nu] ) +  D^R_{[\mu} \tilde Q_{\nu]} + i[\tilde Q_\mu,\tilde Q_\nu]
\] 
where we have introduced the background covariant derivative $D^R_\mu $: $D_\mu^R \phi = \p_\mu \phi + i [W_\mu ,\phi]$.
Let us set $(\p_\mu W_\nu-\p_\nu W_\mu + i[W_\mu,W_\nu] )=V_{\mu\nu}$ .
The action becomes:
\[
S[A]=S[W+\tilde Q]={1\over 4}Tr[F_{\mu\nu}]^2= {1\over 4} Tr[V_{\mu\nu} + D^R_{[\mu}\tilde Q_{\nu]}+i[\tilde Q_\mu,\tilde Q_\nu]]^2
\]
Once the background is treated separately we denote it by:
\[
S[W,\tilde Q]=
{1\over 4}Tr[V_{\mu\nu}]^2 +\hf Tr[V_{\mu\nu}D^{R[\mu}\tilde Q^{\nu]}]+ \hf Tr[V^{\mu\nu} i[\tilde Q_\mu,\tilde Q_\nu]]+
\]
\be
 {1\over 4}Tr[D^R_{[\mu}\tilde Q_{\nu]}]^2 -{1\over 4}Tr[[\tilde Q_\mu,\tilde Q_\nu]]^2 + \hf Tr[D^R_{[\mu}\tilde Q_{\nu]}[\tilde Q^\mu,\tilde Q^\nu]]
\ee
The action is manifestly background gauge covariant - since each term is. The sum of the terms has  the property that it can be expressed as a gauge invariant function of $A$ because $S[W,\tilde Q]=S[A]$. Another way of seeing this is that if we replace $\tilde Q=A-W$, the $W$ dependence cancels out in the sum.  This fact coupled with manifest background gauge invariance
(under which $\delta A_\mu =\p_\mu \Lambda +i[A_\mu ,\Lambda] ;\delta W_\mu = \p_\mu \Lambda + i[W_\mu,\Lambda]$
guarantees that the sum is gauge invariant under (\ref{gauge}).)
Note that
\[
S[W,0] = {1\over 4}Tr[V_{\mu\nu}]^2
\]
is manifestly gauge invariant, because $\tilde Q=0$ is preserved by the gauge transformation, and this sets $A=W$. \footnote{This last
statement is true
for a suitably defined quantum effective action also. This is what makes the background field formalism useful\cite{Abbott}. However we will need only the classical action in this paper.}

Thus the Yang-Mills equation of motion is

\[
D^\mu F_{\mu\nu}=0=D^{R\mu}( V_{\mu\nu}+D^R_{[\mu}\tilde Q_{\nu ]} + i[\tilde Q_\mu,\tilde Q_\nu]) + i[\tilde Q^\mu,( 
V_{\mu\nu}+D^R_{[\mu}\tilde Q_{\nu ]}+ i[\tilde Q_\mu,\tilde Q_\nu])]
\]

These equations are cubic. But we can imagine starting with an action:
\[
{1\over 4}Tr[V_{\mu\nu}]^2 +\hf Tr[V_{\mu\nu}D^{R[\mu}\tilde Q^{\nu]}]+ \hf Tr[V^{\mu\nu} i[\tilde Q_\mu,\tilde Q_\nu]]+
\]
\be
 {1\over 4}Tr[D^R_{[\mu}\tilde Q_{\nu]}]^2 -{1\over 4}Tr[\Phi_{\mu\nu}]^2+{1\over 4}Tr[i[\tilde Q_\mu,\tilde Q_\nu]\Phi^{\mu\nu}]  + \hf Tr[D^R_{[\mu}\tilde Q_{\nu]}[\tilde Q^\mu,\tilde Q^\nu]]
\ee
where $\Phi$ is a very massive mode, an adjoint of $SU(2)$, and we neglect the derivative part of the kinetic term at low energies.
Solving for the $\Phi$ equation would give back the original low energy action. Now the EOM are quadratic:
\[
\Phi_{\mu \nu}= i[\tilde Q_\mu,\tilde Q_\nu]
\]
\[
D^{R\mu}D^R_{[\mu}\tilde Q_{\nu]} + [\tilde Q_\mu, \Phi_{\mu \nu}] + [\tilde Q_\mu, D^R_{[\mu}\tilde Q _{\nu]}] + D^{R\mu}[\tilde Q_\mu,\tilde Q_\nu]+
\]
\[
i[\tilde Q^\mu, V_{\mu\nu}] + D^{R\mu}V_{\mu\nu}=0
\]
Notice also that the equations are background gauge covariant. If we set $\tilde Q = A_\mu - W_\mu$ in the action, we would get back the original Yang-Mills action without $W$. So in principle we could therefore choose $W=0$. However then the first equation becomes
\[
\Phi_{\mu \nu}= i[ A_\mu, A_\nu]
\]
and we do not see any manifest\footnote{Technically, "manifest" here means being able to write equations in terms of fields transforming in  linear representations of the group} background (or other) gauge covariance. The same is true in the second equation. 
Thus without background fields the individual equations do not have any manifest symmetry. Neverthless if we substitute for
$\Phi$ we get the original Yang Mills gauge covariant equation.

The lesson is that the role played by the arbitrary reference field $W$ is to make each equation manifestly covariant under a background  gauge transformation. Thus in the intermediate stages of the calculation, some covariance property is manifest. This guarantees that when $\Phi$ is eliminated by its equation of motion, then the result will continue to be background covariant. Then as we have seen, the property $S[W,\tilde Q]= S[W+\tilde Q]$ guarantees that the result has the original gauge invariance.

In the problem at hand, $A$ is replaced by $h_{\mu \nu}$, $W$ by $h^R$ and $\tilde Q$ by $\tilde h$. We break up our original action
into a kinetic and interaction term. 
\[
S(h)=\int dz~(\eta _{\mu \nu}+h_{\mu \nu}(X)) \p_z X^\mu \p_\zb X^\nu 
\]
\[
 = \int dz~(\eta _{\mu \nu}+h^R_{\mu \nu}(X)) \p_z X^\mu \p_\zb X^\nu+\tilde h_{\mu \nu}(X) \p_z X^\mu \p_\zb X^\nu
\]
where $\tilde h\equiv h-h^R$.
The transformation rules are as given earlier (for infinitesimal $\eps^\mu$, and $\eps_\mu \equiv \eta _{\mu \nu} \eps ^\nu$)
\be	\label{gct}
\delta  h_{\mu \nu}= \eps _{( \mu,\nu)} + \eps ^\la h_{\mu \nu,\la} + \eps ^\la_{~,\mu}h_{\la,\nu}+\eps ^\la_{~,\nu}h_{\mu,\la}
\ee
\be	\label{bgct}
\delta  h^R_{\mu \nu}= \eps _{( \mu,\nu)} + \eps ^\la h^R_{\mu \nu,\la} + \eps ^\la_{~,\mu}h^R_{\la,\nu}+\eps ^\la_{~,\nu}h^R_{\mu,\la}
~~~\delta  \tilde h_{\mu \nu}=  \eps ^\la \tilde h_{\mu \nu,\la} + \eps ^\la_{~,\mu}\tilde h_{\la,\nu}+\eps ^\la_{~,\nu}h^R_{\mu,\la}
\ee
Thus $\tilde h$ (like $\tilde Q$ above) transforms homogeneously.
While the sum of kinetic plus interaction term is coordinate invariant (\ref{gct}), each term individually is invariant only under the background gauge transformation (\ref{bgct}). We expect that $\beta (h) = \beta (h^R+\tilde h)=\beta (h^R,\tilde h)$. This is because both terms correspond to the same beta function. In one we treat $h$ perturbatively to all orders, in the other we treat $h^R$ as a background and $\tilde h$ perturbatively. The sum of the infinite series should add up (formally, i.e. in some region of convergence) to satisfy this equation.

Now we can also calculate the ERG, which is only quadratic in fields, and has all the massive modes. If we solve for the massive modes
(as in the Yang Mills example) we should recover the low energy non polynomial beta function. If we use the background field formalism, each equation is guaranteed to be background gauge covariant. Then the result of solving for the massive modes gives us the low energy $\beta$ function, which also has background gauge covariance. Now using $\beta (h^R+\tilde h) = \beta (h)$ we see
that the result in fact has full covariance. Equivalently, the initial action does not depend on $h^R$ and $h^R$ is completely arbitrary. So if we write the final answer entirely in terms of $h$ and $h^R$, $h^R$ has to drop out of the final result as expressed by (\ref{hR}). Then the background covariance reduces to ordinary covariance.

\end{document}